\documentclass[useAMS,usenatbib]{mnras}
\usepackage{graphicx}
\usepackage{grffile}\usepackage{epsf}
\usepackage{amsmath}
\usepackage{amssymb}
\usepackage{bm}

\usepackage{enumitem}
\setlist[enumerate,1]{leftmargin=*}

\newcommand{\kmsmpc}{\kms\;{\rm Mpc}^{-1}}
\newcommand{\lya}{{\rm Ly}\alpha}

\newcommand{\kms}{{\rm km}\,{\rm s}^{-1}}

\newcommand{\cmc}{{\rm cm}^{-3}}

\newcommand{\Zsolar}{\;{\rm Z}_{\odot}}
\newcommand{\msolar}{{\rm M}_{\odot}}
\newcommand{\msolaryr}{{\rm M}_{\odot} {\rm yr}^{-1}}
\newcommand{\Msolkmskpc}{\rm M_{\odot}\,{\rm km}\,{\rm s}^{-1}\, {\rm kpc}}

\newcommand{\gad}{{\sc Gadget-3}}
\newcommand{\CIV}{\hbox{C\,{\sc iv}}}
\newcommand{\CII}{\hbox{C\,{\sc ii}}}
\newcommand{\MgII}{{\hbox{Mg\,{\sc ii}}}}
\newcommand{\SiII}{{\hbox{Si\,{\sc ii}}}}
\newcommand{\SiIII}{{\hbox{Si\,{\sc iii}}}}
\newcommand{\SiIV}{\hbox{Si\,{\sc iv}}}

\newcommand{\OVI}{\hbox{O\,{\sc vi}}}
\newcommand{\OVII}{\hbox{O\,{\sc vii}}}

\newcommand{\HI}{{\hbox{H\,{\sc i}}}}

\newcommand{\nh}{{n_{\rm H}}}

\hypersetup{draft} %% added by JW because hyperlinks were throwing up a bizarre error. Keep this for draft stage, may have to fix refs hyperlinks later? 

\begin{document}

\title[The changing CGM: physical properties]{The changing circumgalactic medium over the last 10 Gyr I: physical and dynamical properties}

\author[E. Huscher et al.]{
\parbox[t]{\textwidth}{\vspace{-1cm}
Ezra Huscher$^{1}$\thanks{ezra.huscher@colorado.edu}, Benjamin D. Oppenheimer$^{1,2}$, Alice Lonardi$^{1}$, Robert A. Crain$^{3}$, Alexander J. Richings$^{4}$, Joop~Schaye$^{5}$
}\\\\
$^1$CASA, Department of Astrophysical and Planetary Sciences, University of Colorado, 389 UCB, Boulder, CO 80309, USA\\
$^2$Harvard-Smithsonian Center for Astrophysics, 60 Garden St., Cambridge, MA 02138, USA\\
$^3$Astrophysics Research Institute, Liverpool John Moores University, 146 Brownlow Hill, Liverpool, L3 5RF, UK\\
$^4$Institute for Computational Cosmology, Durham University, South Road, Durham, DH1 3LE, UK\\
$^5$Leiden Observatory, Leiden University, P.O. Box 9513, 2300 RA, Leiden, The Netherlands\\
}
\maketitle

\pubyear{2020}

\maketitle

\label{firstpage}

\begin{abstract}

We present an analysis of the physical and dynamical states of two sets of EAGLE zoom simulations of galaxy haloes, one at high redshift ($z=2-3$) and the other at low redshift ($z=0$), with masses of $\approx 10^{12} \msolar$. Our focus is how the circumgalactic medium (CGM) of these $L^*$ star-forming galaxies change over the last 10 Gyr. We find that the high-$z$ CGM is almost equally divided between the ``cool" ($T<10^5$ K) and ``hot'' ($T\geq 10^5$ K) phases, while the low-$z$ hot CGM phase contains $5\times$ more mass. The  high-$z$ hot CGM contains 60\% more  metals than the cool CGM, while the low-$z$ cool CGM contains 35\% more metals than the hot CGM content. The metals are evenly distributed radially between the hot and cool phases throughout the high-$z$ CGM. At high $z$, the CGM volume is dominated by hot outflows, cool gas is mainly inflowing, but cool metals are flowing outward.  At low $z$, the cool metals dominate the interior and the hot metals are more prevalent at larger radii. The low-$z$ cool CGM has tangential motions consistent with rotational support out to $0.2 R_{200}$, often exhibiting $r \approx 40$ kpc disc-like structures.  The low-$z$ hot CGM has several times greater angular momentum than the cool CGM, and a more flattened radial density profile than the high-$z$ hot CGM.  This study verifies that, just as galaxies demonstrate significant evolutionary stages over cosmic time, the gaseous haloes surrounding them also undergo considerable changes of their own both in physical characteristics of density, temperature, and metallicity, and dynamic properties of velocity and angular momentum.

\end{abstract}


\section{Introduction}

It is well established that galaxies are surrounded by gaseous reservoirs of baryons and metals extending far beyond the optical stellar component.  Observational probes of a galaxy's circumgalactic medium (CGM) span across cosmic time from the relatively local Universe \citep[e.g.][]{tumlinson11, stocke13,borthakur15,burchett15,johnson15} to the peak of cosmic star formation \citep[e.g.][]{adelberger03,steidel10,turner14,rudie19} and beyond.  While galaxies change dramatically in appearance from the {\it high-redshift} epoch, sometimes referred to as ``Cosmic Noon" ($z\approx 2-3$), to comparatively nearby {\it low-redshift} galaxies ($z\la 0.3$), it is less well understood how the CGM of these galaxies change.

Galaxies evolve dramatically in appearance from high to low redshift.  If one selects halo masses at both epochs which provide the most efficient conversion of baryons to stars, $M_{\rm halo} \sim 10^{12} \msolar$ \citep{behroozi13b}, the central galaxies predicted to inhabit them are dramatically different. Galaxies at high-$z$ are forming stars in excess of $10\times$ the present day rate \citep{pettini01}.  The galaxies are more compact \citep{vandokkum08} and their morphologies more asymmetric \citep{abraham96}, despite having similar stellar masses \citep{behroozi13a,moster13}.  

Comparing the CGM across 10 Gyr of cosmic time (i.e. from $z=3$ to $z=0$) has not received the same attention as galaxies, but absorption line measurements of the same species do exist at both epochs.  \citet{chen12} compared the UV absorption line probes of the CGM at $z\approx 0.1$ and $z\approx 2.2$ finding that the spatial extent and mean absorption strengths of UV transitions change little over 10 Gyr of evolution around similar mass galaxies. A comparison of column densities of a variety of UV absorption species around star-forming $L^*$ haloes between \citet{werk13} at $z\sim 0.2$ and \citet{rudie19} at $z\approx 2$ finds similar column densities as a function of physical separation (i.e. physical not comoving kpc). 

While comparing the low-$z$ and high-$z$ CGM is now possible due to growing observational databases at both epochs, it is also necessary to use sophisticated theoretical tools to contrast the physical properties of gaseous haloes across time. Cosmological hydrodynamic simulation codes have been developed that reproduce crucial properties of galaxy populations at both low and high redshift, including the EAGLE \citep[Evolution and Assembly of GaLaxies and their Environments,][]{schaye15}, Illustris-TNG \citep{pillepich18}, Horizon-AGN \citep{dubois16}, and SIMBA \citep{dave19} simulations.  These simulation suites apply detailed modules for a variety of processes associated with galaxy formation, including gas cooling, star formation, the growth of supermassive black holes (SMBHs), and stellar and SMBH superwind feedback.  EAGLE is not calibrated to reproduce observations of gas \citep{crain15}, but other suites have been calibrated against some gas observations \citep[e.g.][]{pillepich18}.  Therefore, EAGLE provides genuine predictions for the physical and observational characteristics of the CGM.  

In this paper, we expand upon a set of high-resolution, cosmological zoom simulations called the EAGLE-CGM Project introduced by \citet[][hereafter O16]{opp16} using the EAGLE prescription.  These simulations ran at higher resolution than the main EAGLE volume and integrated the CHIMES non-equilibrium chemistry and cooling module developed in \citet{opp13a} and \cite{richings14a}.  We complement the original set of zoom simulations reaching typical halo masses of $M_{200}=10^{12} \msolar$ by the $z\la 0.3$ epoch with a set of high-$z$ ($z=2-3$) galaxy haloes also with $M_{200}=10^{12} \msolar$, where $M_{200}$ is the mass enclosed within a sphere with mean density of $200\times$ the critical density. These high-$z$ haloes are the progenitors of low-redshift EAGLE-CGM zooms of $M_{200}\sim 10^{13} \msolar$ haloes, often hosting passive galaxies at $z\leq 0.2$.  

This paper is the first of a series of papers investigating the CGM at similar halo mass across different epochs. Here, we compare the physical and kinematic properties of gaseous haloes hosting galaxies that are most efficient at turning their baryons into stars. At low-redshift, these are disc galaxies, often with ``grand design'' spiral morphologies, with typical $M_*= 1-3\times 10^{10} \msolar$ and ${\rm SFR} = 0.5-3.0 \msolaryr$.  At high-redshift, these galaxies have similar stellar masses, $M_*= 0.6-3\times 10^{10} \msolar$ but with ${\rm SFR} = 5-45 \msolaryr$, which are consistent with the properties of Lyman-Break Galaxies \citep{steidel96}.  

We aim to compare and contrast several fundamental properties of star-forming galaxies occupying $10^{12} \msolar$ haloes across cosmic time.  These include the gaseous mass budget within haloes, divided into categories of cool (defined here at $T<10^5$ K), hot ($T>10^5$ K), and interstellar (defined as star-forming) in \S\ref{sec:mass}.  We also consider the metal contents and metallicities of gaseous haloes in \S\ref{sec:metals}.  The velocities of cool and hot CGM components are compared across epochs in \S\ref{sec:velocity}.  In \S\ref{sec:angularmom}, we compare the angular momenta of the CGM components, including the relative angles of their axes.  Lastly, we present hot gas radial profiles in \S\ref{sec:hotgas}. 

We emphasize that the low-$z$ EAGLE-CGM haloes have been well-tested against observational datasets of UV ions.  \citet{opp18c} found good agreement for a number of low metal ions observed by COS-Halos \citep{werk13}, including $\CII$, $\SiII$, $\SiIII$, and $\SiIV$, but under-produced the observed $\MgII$ strengths.  O16 reproduced the observed correlation between $\OVI$ and sSFR \citep{tumlinson11}, but under-predicted their column densities, which \citet{opp18a} later argued could be enhanced to observed levels by flickering AGN flash-ionizing the CGM and leaving metals over-ionized long after the AGN turns off.  In a companion paper (Lonardi et al., in prep), we will show that the high-$z$ zoom haloes broadly reproduce the observed column densities of \citet{rudie19}.  This series of papers contrasting the two selected epochs relies on testing our simulations against observations using ion-by-ion tracking of the non-equilibrium module \citep{opp13a,richings14a}.  

The paper is organized as follows.  In \S\ref{sec:methods}, we review the code used for EAGLE-CGM simulations and introduce our set of simulations.  The main results are presented in \S\ref{sec:results} on topics of CGM mass (\S\ref{sec:mass}), metals (\S\ref{sec:metals}), velocities (\S\ref{sec:velocity}), and angular momentum (\S\ref{sec:angularmom}), as well as hot gas profiles (\S\ref{sec:hotgas}).  We discuss several findings in detail in \S\ref{sec:discussion}.  We summarize in \S\ref{sec:summary}.  Physical kpc units are used throughout.

%Cosmological simulations help explain observations within a mathematical framework of known physics. While the computation involved to develop massive primordial clouds into distinct galaxies is considerable, once complete we attain dozens or even hundreds of examples of galactic evolution bred from identical initial conditions ready for statistical analysis.

%The development of this field is progressing rapidly alongside improving computing power, allowing for detailed simulations to follow the actions of smaller and smaller groups of particles. Here we use hydrodynamic simulations from the EAGLE (Evolution and Assembly of GaLaxies and their Environments) project \citep{schaye15} to compare the physical characteristics of galaxy haloes at differing redshifts.

%Cite with parentheses around it \citep{opp16}
%Cite as part of Text, e.g. \citet{opp16} showed that.. ]

\begin{table*}
\caption{Zoom simulations}
\begin{tabular}{lrrrrrclrrrrrc}
\hline
\multicolumn{7}{|c|}{Mass sums within $R_{200}$ for high-$z$ haloes}& \multicolumn{7}{|c|}{Mass sums within $R_{200}$ for low-$z$ haloes}\\
\hline
     Halo & $z$ & $M_{\rm cool}^{1}$ & $M_{\rm hot}^{1}$ & $M_{\rm CGM}^{1}$ & $M_{\rm 200}^{1}$ & $\frac{M_{\rm CGM}}{M_{\rm 200}} \frac{\Omega_{m}}{\Omega_{b}}^{2}$  &
     Halo & $z$ & $M_{\rm cool}^{1}$ & $M_{\rm hot}^{1}$ & $M_{\rm CGM}^{1}$ & $M_{\rm 200}^{1}$ & $\frac{M_{\rm CGM}}{M_{\rm 200}} \frac{\Omega_{m}}{\Omega_{b}}^{2}$ \\
\hline
     HiZ000& 3.02 &1.79&2.49&4.28& 115&0.24  &   LoZ001&0& 1.47  & 6.89&8.36&129& 0.41  \\
     HiZ002& 2.24 &2.94&2.43&5.37& 100&0.34  &   LoZ002&0& 0.57  & 11.50&12.10&191& 0.40  \\
     HiZ003& 2.24 &3.50&3.90&7.40& 96&0.49  &   LoZ003&0& 2.21  & 9.61&11.80&151& 0.50 \\
     HiZ004& 3.02 &2.76&4.90&7.67& 151&0.32  &   LoZ004&0& 1.25  & 5.07&6.32&105& 0.38 \\
     HiZ005& 2.01 &2.86&3.66&6.52& 110&0.38  &   LoZ005&0& 2.34  & 15.20&17.50&170& 0.66  \\
     HiZ006& 3.02 &1.63&1.40&3.03& 79&0.24  &   LoZ006&0& 0.54  & 4.73&5.26&87& 0.38  \\
     HiZ007& 2.24 &2.58 &2.29 &4.87&83 &0.37  & LoZ007&0& 2.97  & 4.85&7.81&71& 0.70 \\
     HiZ008& 3.02 &2.38&4.05&6.43& 145&0.28  &   LoZ008&0 & 0.93  & 3.89&4.83&72& 0.42\\
     HiZ009& 3.02 &2.78&4.24&7.02& 115&0.39  &   LoZ009&0& 0.65  & 3.47&4.11&76& 0.34  \\
     
     Averages:& 2.63 &2.58&3.26&5.84& 110&0.34  &   Averages:&0& 1.44  & 7.25&8.68&117& 0.47  \\
     
\hline
\end{tabular}
\\
\parbox{25cm}{
$^1$ $M_{\rm cool}$ ($M_{\rm hot}$) includes CGM gas at $<10^5$ ($\geq 10^5$) K.  $M_{\rm CGM}=M_{\rm cool}+M_{\rm hot}$.  In units of $10^{10} M_{\odot}$.\\
$^2$ The fraction of a halo's cosmic proportion of baryons residing in the CGM inside $R_{200}$.%, where $R_{200}$ is the radius of a sphere with mean density of $200\times$ the critical density.  
}
\label{tab:masses}
\end{table*}

\section{Methods} \label{sec:methods}

%In general, smooth particle hydrodynamic (SPH) code solves the Euler equations for conservation of mass, momentum, and energy in fluids. The N-body+SPH code Gadget3 (Springel 2005) uses a tree algorithm to follow the progress of particles in order to simulate the formation of galaxies in a $\lambda$ cold dark matter universe. The Virgo consortium improved Gadget by developing subgrid models to better accommodate existent phenomena, such as radiative cooling, feedback from both star formation and active galactic nuclei, and gas accretion on supermassive black holes.

\subsection{The EAGLE simulation code} 

We introduce the simulations in this section, and refer the reader to \S2 of O16 for further details.  We employ the EAGLE hydrodynamic simulation code introduced in \citet{schaye15}, which uses the \gad~N-body+SPH code \citep[see][]{springel05}, plus extensive modifications to simulate galaxy formation described below.  The \citet{planck13} cosmological parameters are adopted: $\Omega_{\rm m}=0.307$, $\Omega_{\Lambda}=0.693$, $\Omega_b=0.04825$, $H_0= 67.77$ $\kmsmpc$, $\sigma_8=0.8288$, and $n_{\rm s}=0.9611$.  EAGLE applies the \citet{hopkins13} pressure-entropy SPH formulation using a C2
\citet{wendland95} 58-neighbour kernel along with several other
hydrodynamic modifications collectively referred to as ``Anarchy''
\citep[Appendix A of Schaye et al. 2015 and ][]{schaller15}.  

The EAGLE simulations include subgrid prescriptions for radiative cooling \citep{wiersma09a}, star formation \citep{schaye08}, stellar evolution and chemical enrichment \citep{wiersma09b}, and superwind feedback associated with star formation \citep{dalla12} and black hole growth \citep{rosas16,schaye15}. The parameters governing the efficiency of the star formation feedback were calibrated to reproduce the present-day stellar masses of galaxies, whilst also recovering galaxy discs with realistic sizes. Those governing feedback associated with black hole growth were calibrated to reproduce the present-day relationship between the stellar mass of galaxies and the mass of their central black holes. The feedback calibration strategy is discussed in detail by \citet{crain15}.

\subsection{EAGLE zoom simulations}

We analyse two samples of haloes, one at high $z$ with redshifts ranging from $z=2.24$ to 3.02 at halo masses from $M_{200} = 10^{11.90}$ to $10^{12.18} \msolar$, and one at $z=0$ with $M_{200} = 10^{11.85}$ to $10^{12.28} \msolar$.  We list these haloes in Table \ref{tab:masses}, and use identifiers of ``HiZ00X" and ``LoZ00X" for the individual haloes, where X is the halo number.  The ``LoZ00X" haloes are the same haloes listed in Table 1 of O16 as ``Gal00X", but with values listed at $z=0$ as opposed to $z=0.205$ in that paper.  The ``HiZ00X" haloes are virialized haloes selected from the ``Grp00X" zooms listed in the same O16 Table, but selected to have a $M_{200}\sim 10^{12} \msolar$ halo at high $z$.  Hence we are comparing to virialized progenitors of $M_{200} \sim 10^{13} \msolar$ $z=0$ haloes. 

The ``LoZ" haloes are selected from the EAGLE Recal-L025N0752 simulation, and rerun with the CHIMES non-equilibrium (NEQ) ionization and cooling module for diffuse gas described in \citet{opp13a} and implemented in \citet{richings14a} starting at $z=0.503$.  These are identical to the runs listed in O16, and have the {\it M5.3} resolution of O16 corresponding to an SPH particle mass  $m_{\rm SPH} = 2.2\times
10^5 \msolar$, using the notation {\it M}[log($m_{\rm SPH}/\msolar$)].  O16 demonstrated that these low-$z$ haloes follow the $M_*$ and sSFR relations of the Recal-L025N0752 simulation, and argued that these haloes are generalizable to the larger population of haloes hosting $L^*$ star-forming galaxies in this 25$^3$ Mpc$^3$ simulation.  

The ``HiZ" haloes were originally selected from the EAGLE Ref-L100N1504, $100^3$ Mpc$^{3}$ volume and also use the {\it M5.3} resolution.  However, we re-ran all of these simulations using the NEQ module beginning at $z=4$ to follow the haloes to at least the redshift listed in Table \ref{tab:masses}.  We describe the NEQ module in Lonardi et al. (in prep) when we present CGM ion column densities, but note here that O16 found no significant differences in physical or dynamical halo properties compared to runs evolved with cooling rates in chemical equilibrium.  All simulations in this paper use a Plummer-equivalent softening length of 350 proper pc at $z<2.8$, and 1.33 comoving kpc at $z>2.8$.  All zooms have the same resolution as the EAGLE Recal-L025N0752 run.  There does not exist a statistical sample of high-$z$ $\approx 10^{12} \msolar$ haloes in the Recal-L025N0752 volume to test how representative these haloes are, but O16 did argue that their $z=0$ descendants exhibit typical galaxy properties compared to the lower resolution Ref-L100N1504 simulation.

%We handpicked 18 haloes from a larger EAGLE simulation run in a 10Mpc box. Each halo was centered and rotated for a face-on view, and then exported as a snapshot for closer study. All 18 haloes fit the following criteria:
%\begin{enumerate}
%  \item Comparable mass to the Milky Way ($\sim 10^{12} M_{\odot}$)
%  \item No satellite haloes in the simulation box
%  \item No obvious artifacts or irregularities
%\end{enumerate}
%\noindent

\subsection{Definition of temperature and ISM phases} \label{sec:phasedef}

\begin{figure}
    \includegraphics[width=0.49\textwidth]{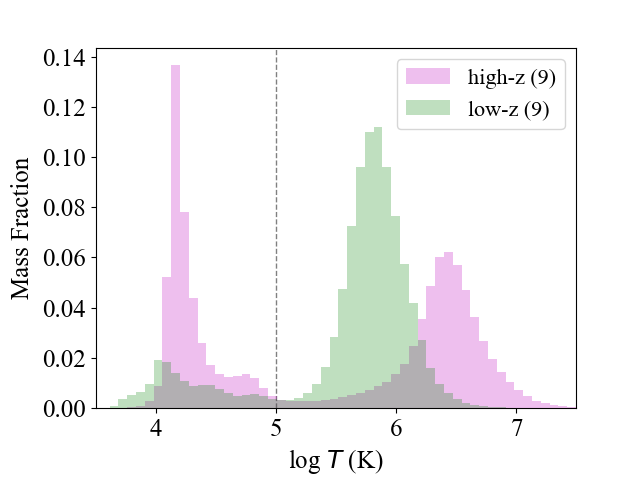}
    \caption{ Normalized temperature distribution of CGM gas out to $R_{200}$ in all nine high-z haloes (pink) and all nine low-z haloes (green). The temperature division we have chosen to separate hot and cool gas ($10^5$ K) is denoted with a dotted line. }
    \label{fig:masshist}
\end{figure}

Throughout we divide gas into ``cool" and ``hot" phases using a temperature cut of $10^5$ K.  Often gas in the $T=10^{5-7}$ K range is considered ``warm-hot", with ``hot" being reserved for $>10^6$ K gas.  The main reason we use a $10^5$ K cut is because it divides cool gas, which is often in thermal equilibrium with the metagalactic UV background, from gas that is often heated to the virial temperature of the halo, which is $T_{\rm vir}\approx 10^6$ K for our haloes as shown in Figure \ref{fig:masshist}.  \citet{correa18} explored the cooling properties of halo gas in the main EAGLE 100 Mpc simulation, and also found that $T=10^5$ K represents a clear division between cool and hot gas in $10^{12} \msolar$ haloes with little gas around $10^5$ K indicating that the cool-hot division is relatively insensitive to the precise temperature cut.  We apply this cut additionally because UV photo-ionized absorption lines correspond to gas at $T\la 10^5$ K \citep[e.g.][]{ford13,rahmati16}.  

We define the ISM as any gas with either 1) non-zero SFR or 2) a gas density threshold greater than $\nh=10^{-1} \cmc$.  This specific definition is meant to exclude the cool ISM in the first case and neutral ISM in the second case.  Using only the first criterion results in insignificant changes to CGM masses.  The high-$z$, cool CGM metallicity profile we show in Fig. \ref{fig:metallicity} is most sensitive to the ISM criteria.  Adding the density threshold as a second criterion raises the cool, high-$z$ CGM metallicity inside $0.3 R_{200}$ by a factor of up to two versus the SFR-only criterion, where $R_{200}$ is the radius of the sphere containing $M_{200}$.  This owes to gas with $\nh>10^{-1} \cmc$ being metal-poor and below the EAGLE star formation density criterion. EAGLE's star formation threshold is meant to simulate the transition from atomic to molecular phases, and hence the star formation density threshold increases with decreasing metallicity.

\begin{figure*}
    \includegraphics[width=0.45\linewidth]{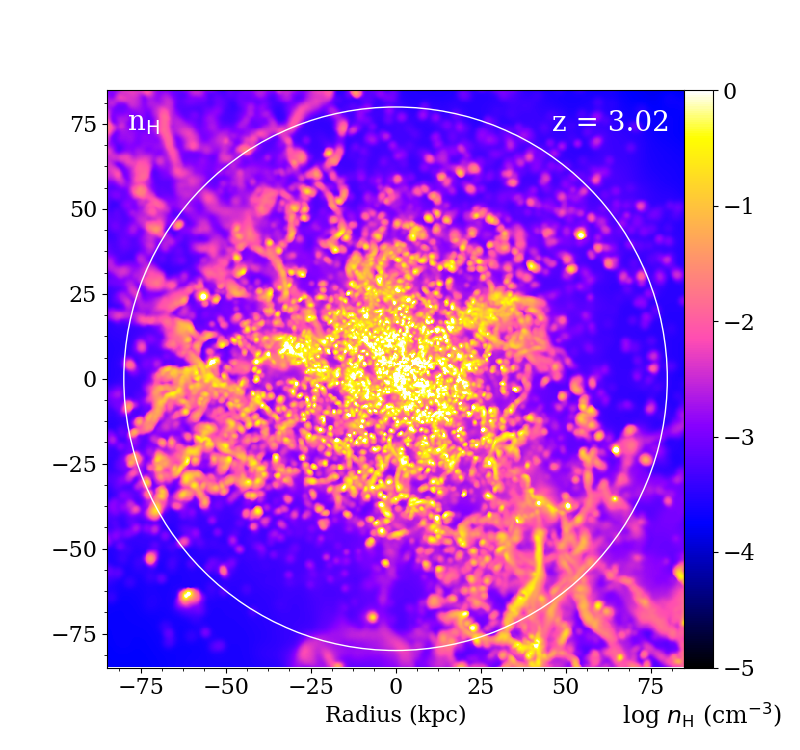}
    \includegraphics[width=0.45\linewidth]{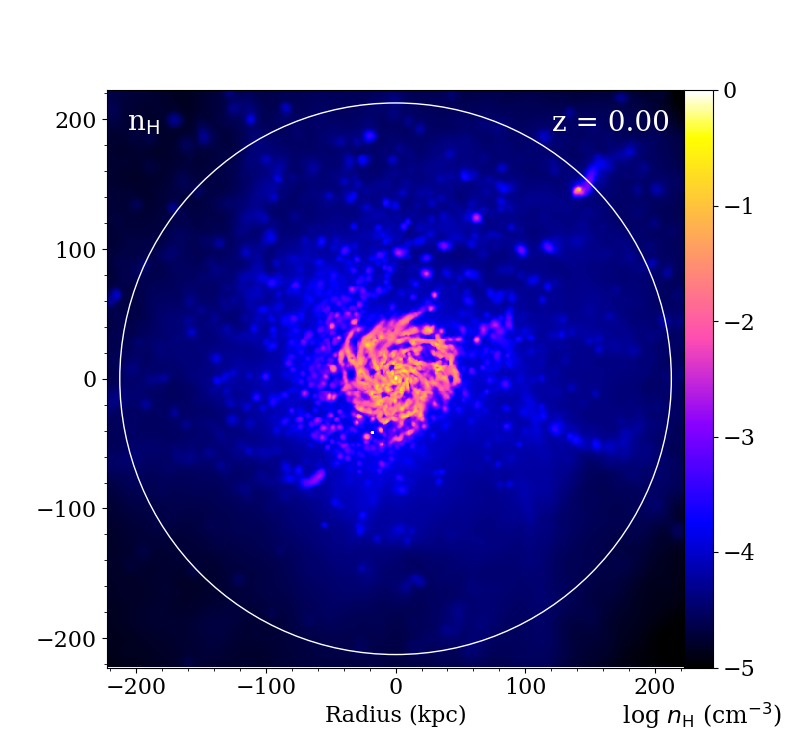}

    \includegraphics[width=0.45\linewidth]{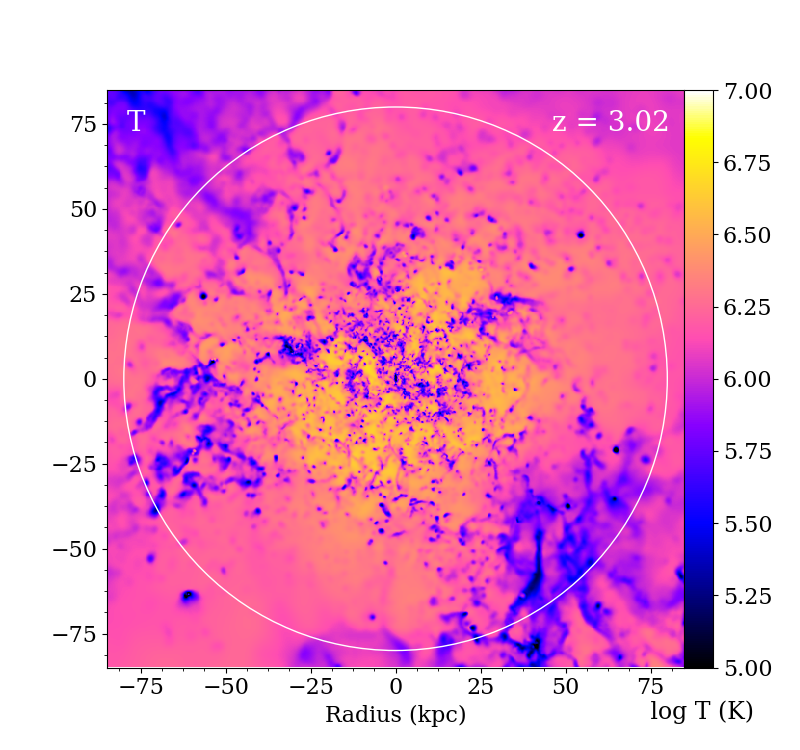}
    \includegraphics[width=0.45\linewidth]{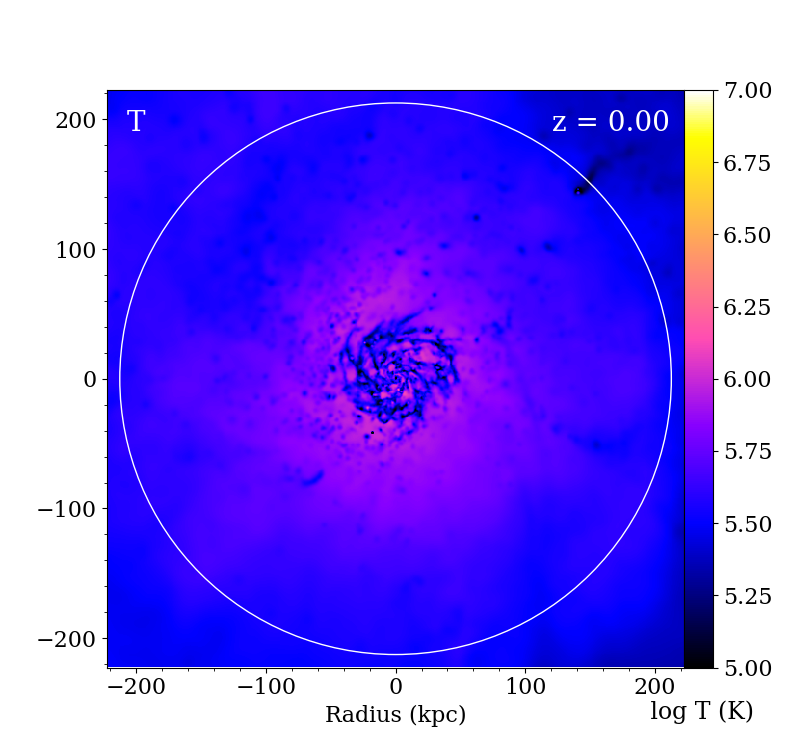}

    \includegraphics[width=0.45\linewidth]{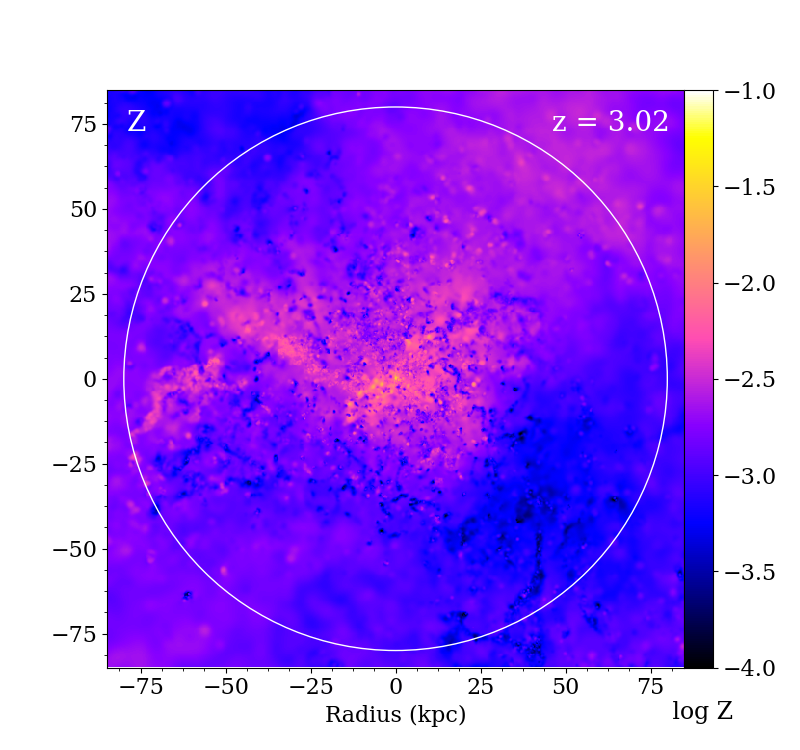} 
    \includegraphics[width=0.45\linewidth]{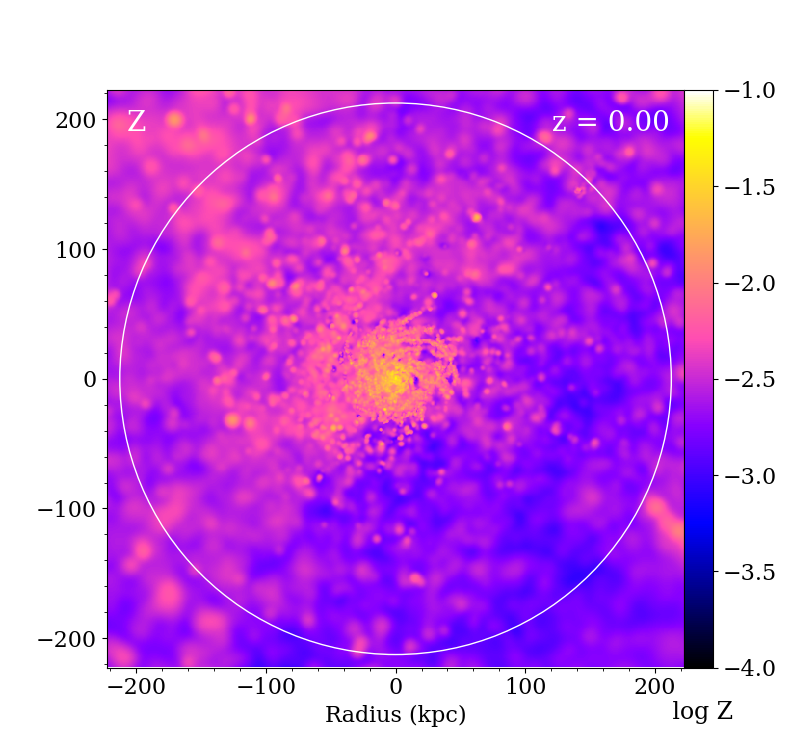}
    \caption{The physical properties of two $10^{12} \msolar$ haloes, HiZ009 on left and LoZ004 on right, plotted out to $R_{200}$, indicated by the white circles. From top to bottom, the panels show hydrogen number density, temperature, and absolute metallicity.}
\label{fig:maps}
\end{figure*}

\section{Physical properties of haloes} \label{sec:results}

 We begin this section by showing two representative $10^{12}\msolar$ haloes in Figure \ref{fig:maps}, one at high $z$ (HiZ009; left) and the other at low $z$ (LoZ004; right) in density ($n_{\mathrm{H}}$), temperature ($T$), and metallicity ($Z$) (from upper to lower panels).  Both haloes are normalized to the virial radius, which is physically $2.5\times$ smaller for the $z=3.02$ halo (213 versus 80 kpc).  More dense, cool gas extends throughout the halo at high $z$, while this gas organizes itself into a disc structure extending nearly 100 kpc across at $z=0$.  At high $z$, the cool, dense gas is less metal enriched, which suggests that it is associated with cold accretion flows \citep[e.g.][]{keres05, dekel06, keres09, vandevoort12}.  At low $z$, the cool, dense gas appears more metal enriched in the disc structure.  The volume-filling, hot gas medium is hotter at high $z$ than at low $z$, which is to be expected given that the virial temperature scales with $R_{200}^{-1}$ at fixed mass, resulting in a temperature scale $\sim 2.5\times$ higher at $z=3$   However, we also see that the hotter gas is more often coincident with metal enrichment at high $z$, which suggests hot, enriched outflows are more common throughout the high-$z$ CGM.  These visual trends prelude the quantitative results we demonstrate in the following subsections.  

\begin{figure*}
    \includegraphics[width=0.49\textwidth]{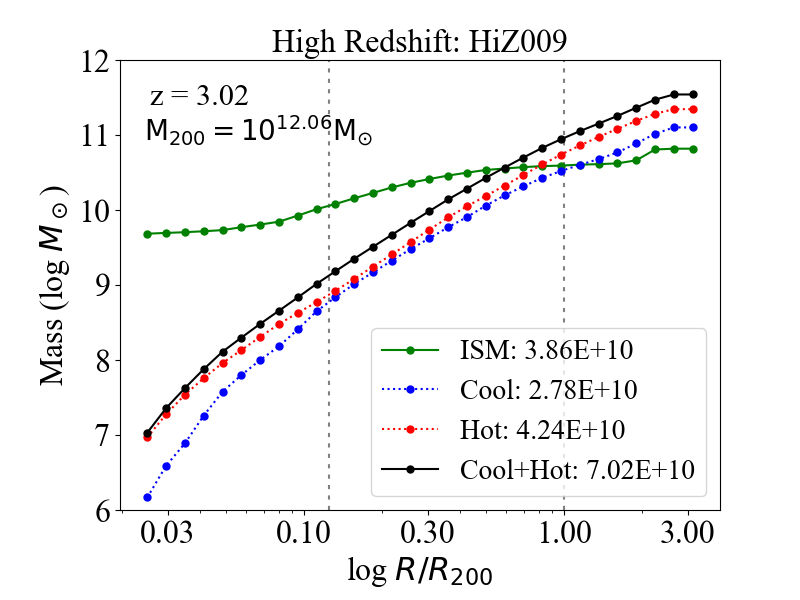}
    \includegraphics[width=0.49\textwidth]{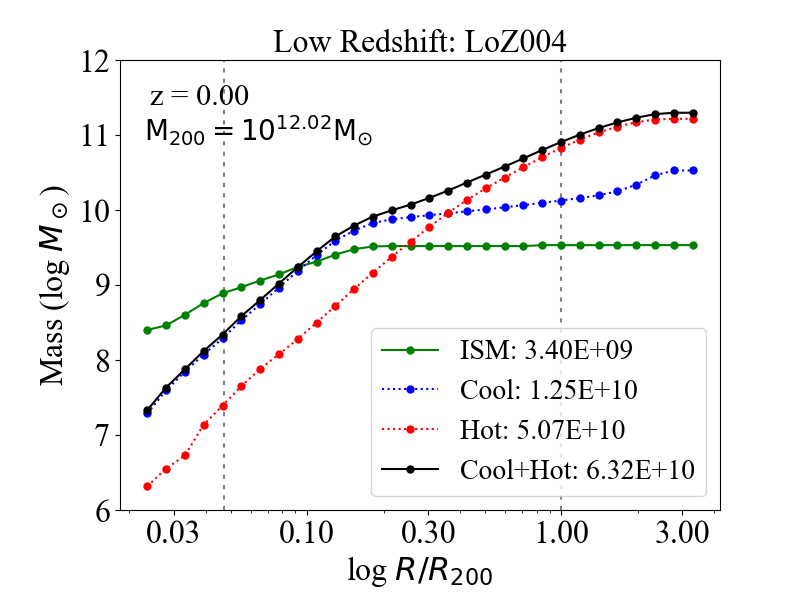}
    \caption{ Cumulative mass as a function of normalized radius for gaseous components in two $M_{200}=10^{12} \msolar$ haloes, one at $z=3$ (HiZ009; left) and one at $z=0$ (LoZ004; right). ISM gas is green, cool CGM gas ($T<10^5$ K) is blue, hot CGM gas ($T\geq 10^5$ K) is red, and the total sum is black. Legends show total masses out to $R_{200}$. Two vertical dashed lines indicate 10 physical kpc and $R_{200}$. The ratio of hot to cool CGM is much higher at low $z$ by $R_{200}$, though the cool phase dominates in the interior of low-$z$ haloes.  The ISM gas is a much smaller fraction of the low-$z$ total halo gas content than at high $z$. Note that a few neighboring galaxies are found in these simulations, as seen here in the ISM increase of HiZ009 at $2R_{200}$.}
\label{fig:massradial}
\end{figure*}

\subsection{Mass} \label{sec:mass}

Figure \ref{fig:massradial} plots the cumulative gas masses for our representative high-$z$ and low-$z$ haloes as a function of galactocentric radius $R$ normalized by $R_{200}$.  The integrated gas masses within $R_{200}$ are listed in the legend.  The green lines indicate the cumulative ISM gas mass, which is $11\times$ higher for the high-$z$ halo ($3.9\times10^{10}\ \msolar$ versus $3.4\times 10^{9}\ \msolar$), and is reflected in the SFR of these galaxies being $30\times$ different ($23.5\ \msolaryr$ versus $0.8\ \msolaryr$). The total CGM masses represented by black lines are more similar between the two haloes ($7.0$ versus $6.3 \times 10^{10}\ \msolar$), but the division between the cool (blue lines) and hot (red lines) phases is rather different.  Cool and hot phases nearly balance each other throughout the CGM at high $z$, but by low $z$ the inner CGM is dominated by cool gas and the outer CGM by hot gas, which becomes the dominant phase throughout the low-$z$ CGM out to $R_{200}$.  

\begin{figure}
    \includegraphics[width=0.49\textwidth]{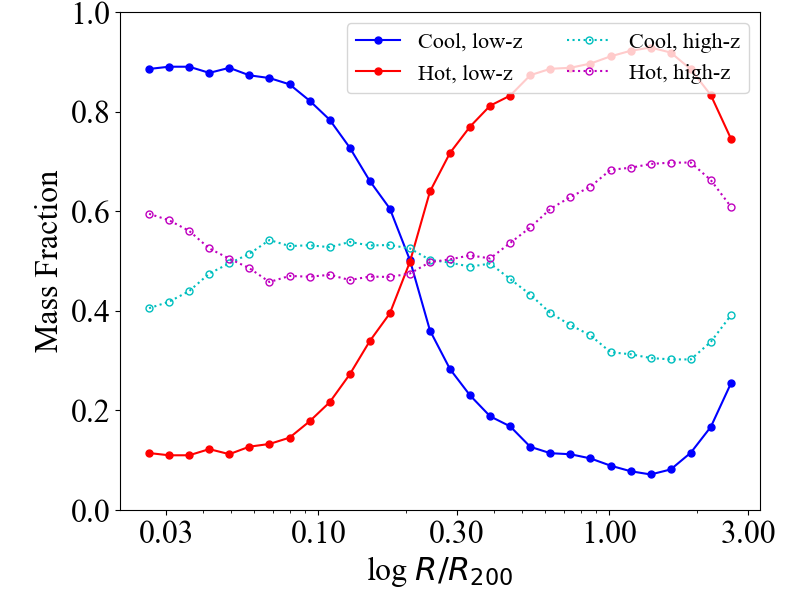}
    \caption{Average differential mass fractions across the 9 high-$z$ (dotted) and 9 low-$z$ haloes (solid), divided into hot and cool gas at $10^5$K. At high redshift, cyan (magenta) indicates cool (hot) gas. At low redshift, blue (red) indicates cool (hot) gas.  Cool (hot) gas dominates in the interior (exterior) of low-$z$ haloes.  High-$z$ haloes have a much more even balance of cool and hot phases throughout the CGM.  The turnover at $\approx 2 R_{200}$ and beyond indicates more cool gas in structures extending outside the CGM and transitioning to the IGM.  }
    \label{fig:masscomp}
\end{figure}

We generalize these results in Figure \ref{fig:masscomp}, which shows the differential division between the cool and hot CGM across each sample of 9 high-$z$ and low-$z$ haloes.  Cool gas accounts for the majority of the inner CGM of low-$z$ haloes, but rapidly transitions to hot gas dominating at larger radii (and hot gas dominating the cumulative CGM mass out to $R_{200}$).  In contrast, the high-$z$ CGM appears less sorted by temperature phases, but retains a similar progression of cool gas being more dominant in the interior.  The ``balance point" where the cool and hot CGM masses at a radius equal each other is $0.5 R_{200}$ at high $z$ and $0.2 R_{200}$ at low $z$.  Beyond $R_{200}$, the cool phase makes a comeback as the extended IGM gas is cooler.  

%The inner CGM is primarily cool gas, giving way to primarily hot gas further out in the halo. The difference in more evolved low-z galaxies is this transition from cool to hot happens closer in. The mass fraction (figure 3) crosses the 50 percent threshold near $0.25R_{200}$ at z=0, where the high redshift transition occurs twice as far out near $0.5R_{200}$. Also, the mass fraction at low redshift is wider throughout most of the halo, reaching a maximum ratio of 90/10 hot to cool gas. High redshift haloes show a more even mix of hot and cool gas, with the greatest ratio at 70/30 hot to cool (Figure xx).

%\begin{figure}[H]
 %   \includegraphics[width=\linewidth]{Plots/combined_highz_means.png}
 %   \caption{Means of High Z haloes. Top: Density; Middle: Metallicity; Bottom: Mass}
%\end{figure}

\subsection{Metals} \label{sec:metals}

\begin{figure*}%[H]
    \includegraphics[width=0.49\textwidth]{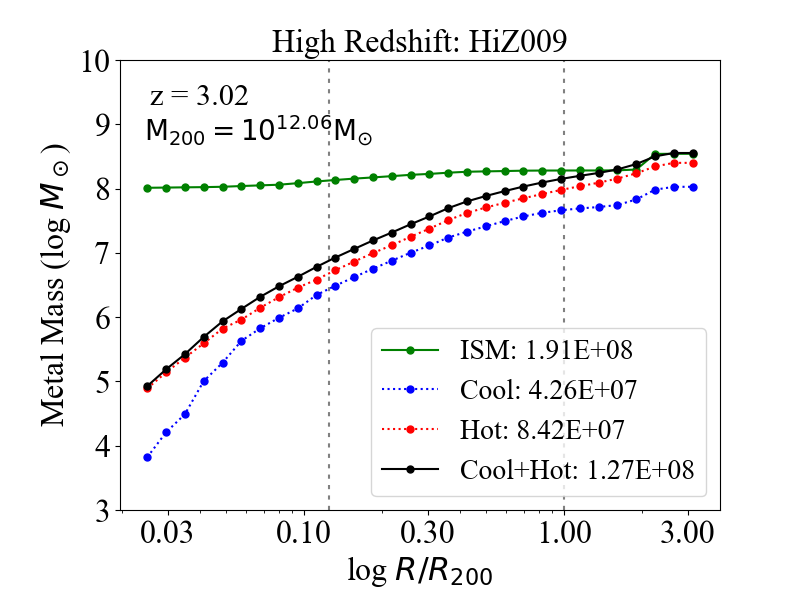}
    \includegraphics[width=0.49\textwidth]{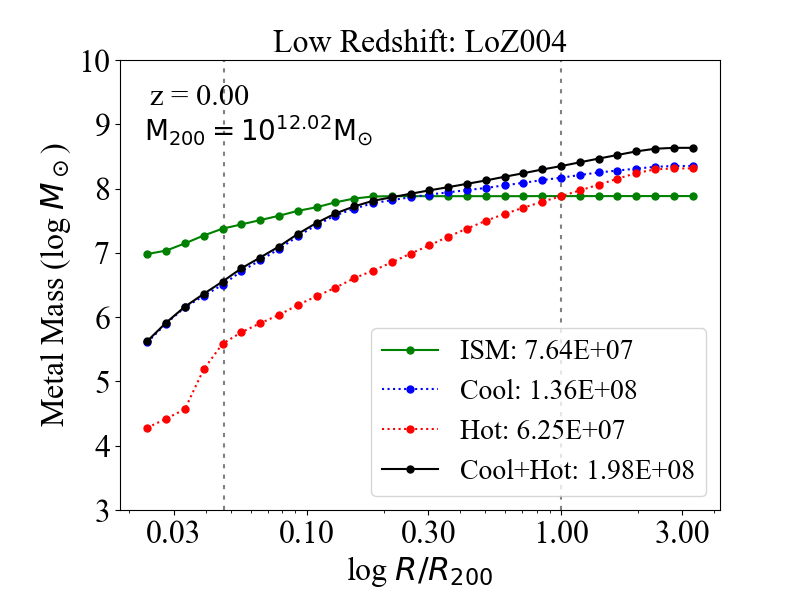}
    \caption{The cumulative gas metal mass plotted as in Fig. \ref{fig:massradial} for the same two haloes, also shown in Fig. \ref{fig:maps}. Legends show mass of metals out to $R_{200}$. At high $z$, nearly all metals are in the ISM out to $R_{200}$. At low $z$, ISM metals dominate only out to $\approx 0.2R_{200}$, where cool metals overtake them. The metal content in the CGM is predominately hot ($T \geq 10^{5}$ K) at higher redshift and predominately cool ($T < 10^{5}$ K) at low redshift.}
    \label{fig:metalradial}
\end{figure*}

\begin{figure}
    \includegraphics[width=0.49\textwidth]{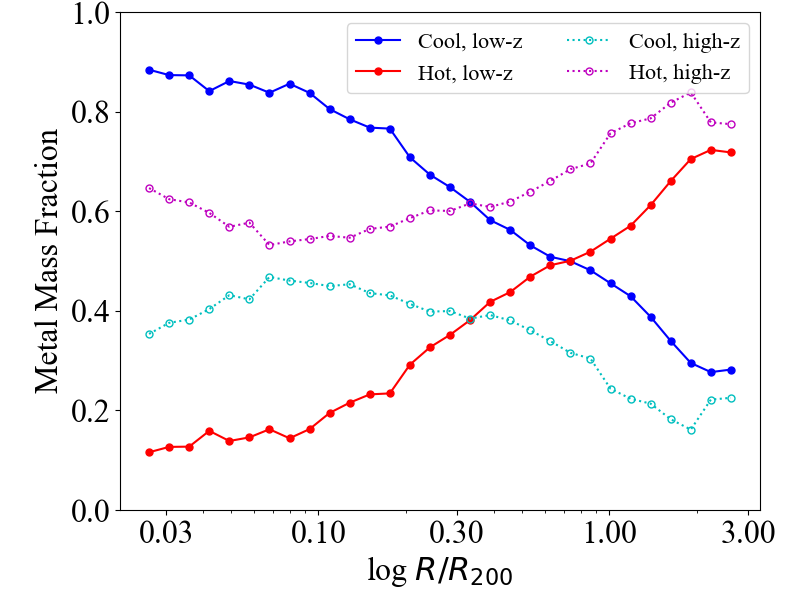}
    \caption{Differential CGM metal fractions divided between cool ($T < 10^{5}$ K) and hot ($T \geq 10^{5}$ K) gas as in Fig. \ref{fig:masscomp}.  Cool metals dominate at $<0.5 R_{200}$ at low $z$, and hot metals dominate beyond $R_{200}$ to at least $3 R_{200}$.  At high $z$, CGM metals are mostly found in the hot phase throughout the CGM and into the IGM.} 
    \label{fig:metalcomp}
\end{figure}

We now turn our attention to the gaseous metal content of haloes.  In Figure \ref{fig:metalradial}, we plot the cumulative metallicity analogous to the mass accumulation of Fig. \ref{fig:massradial} for our representative haloes. At high $z$, we see hot metals in greater abundance than cool metals at every point in the halo all the way out to $3 R_{200}$. In the next subsection, we will show that superwind feedback is pushing out metals in strong, hot outflows at high $z$.  In contrast, cool metals dominate out to $0.5 R_{200}$ in low-$z$ haloes, and then give way to hotter metals beyond $0.5 R_{200}$ when plotting differential fractions on a linear scale in Figure \ref{fig:metalcomp}.  

By $R_{200}$, we see that the ISM holds nearly 50\% more metals than the galaxy's halo at high $z$ (cf. green and black lines in left panel of Fig. \ref{fig:metalradial}). At low $z$, the CGM metals overcome the ISM metal content by $0.6 R_{200}$ and have over $2.5\times$ the ISM metal content by $R_{200}$.  
%[XXX- ROB SUGGESTED BRINGING UP TRAVEL TIME EFFECT, AND QUOTING METAL-WEIGHTED REDSHIFTS.  BDO BELIEVES THIS IS TOO INVOLVED, BUT IT MAY BE HELPFUL TO ASK HOW DID TWO HALOS WITH SIMILAR STELLAR MASSES AND WITHIN A FACTOR OF 2 OF TOTAL METALS PRODUCED HAVE SUCH DIFFERENT PROPORTIONS FOR GASEOUS METALS.  BUT IT IS WORTH PUTTING SOMETHING IN HERE IN TERMS OF METAL-LOADING AND PERHAPS ONE OF PETER MITCHELL'S PAPERS AS TO WHY THE ISM HAS SO MANY MORE METALS THAN THE CGM AT HIGH-Z.  THIS IS FRANKLY SURPRISING BASED ON MY INTUITION, I.E. MASS-LOADING IS HIGHER AT HIGH-Z, AND MZR IS Z ~ YIELD/(1+ETA), BUT WE ARE DEALING WITH ABSOLUTE METAL MASSES AND NOT ISM METALLICITIES, SO THIS IS A BIT OF A RED HERRING] 

\begin{figure*}%[H]
    \includegraphics[width=0.49\textwidth]{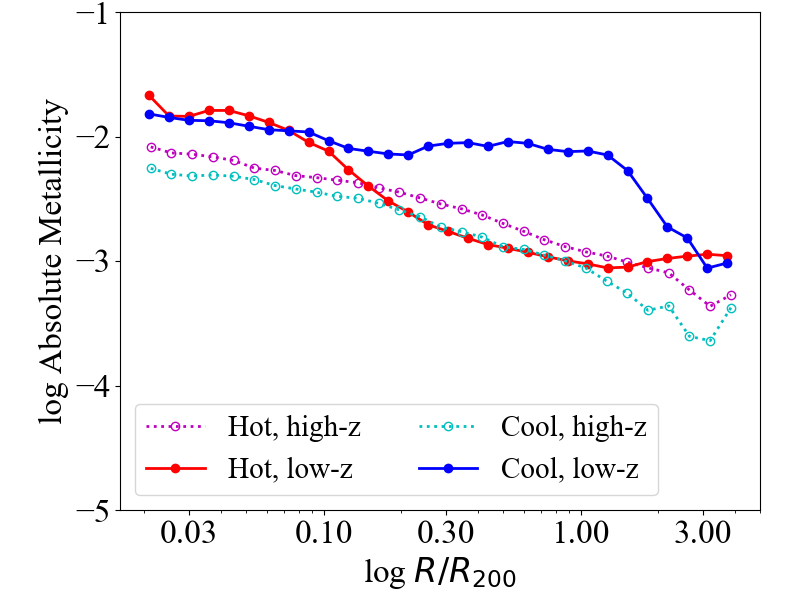}
    \includegraphics[width=0.49\textwidth]{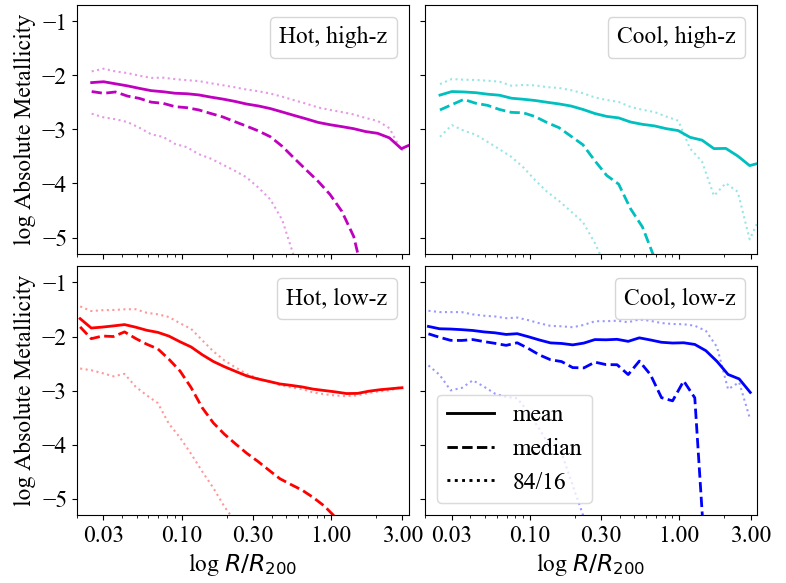}
    \caption{{\it Left:} Mean absolute metallicities at low and high redshift, split into hot ($T \geq 10^{5}$ K) and cool ($T < 10^{5}$ K) gas shown in solid lines. {\it Right:} Individual profiles including dashed lines showing medians and dotted lines showing 16th and 84th percentiles.  Cool gas becomes more enriched by low $z$, and hot gas has similar metallicities in the outer CGM.  The much lower medians for extended high-$z$ cool gas indicate more pristine cool gas in contrast to low-$z$.}
\label{fig:metallicity}
\end{figure*}

In Figure \ref{fig:metallicity}, we plot the absolute metallicity ($Z$) of the CGM phases.  Starting with cool metals, we see a $2-5\times$ greater mean $Z$ at low $z$ compared to high $z$, with a separation that grows at larger radii (left panel). While the average low-$z$ metallicity approaches solar in the interior \citep[][$\Zsolar\equiv 10^{-1.87}$]{asplund09}, the high $z$ metallicity drops below $0.1 \Zsolar$ at $R>0.5 R_{200}$.  The median (dashed lines in right panels) is comparatively much lower at high $z$, with at least half the gas remaining at $Z<10^{-5}$ at $>0.6 R_{200}$.  Hence, most of the outer cool CGM has $Z \la 10^{-3} \Zsolar$, indicating a primordial origin for much of the extended cool CGM.  

The hot phase demonstrates a remarkable contrast to the cool metal evolution, with average metallicities being slightly lower at low $z$ than at high $z$ between $0.15-1.5 R_{200}$.  This extended profile of hot metals is indicative of outflowing thermal winds which are steadily enriching the CGM at high-$z$ as we explore in \S\ref{sec:velocity}.  The low-$z$ inner hot metals approach $\Zsolar$, and their high median and comparatively small dispersions (dotted lines, right panels) indicate widespread enrichment in the interior $0.1 R_{200}$.  Extended low-$z$ haloes have a greater dispersion of metals than at high $z$ (cf. dotted lines at $0.3 R_{200}$), suggesting that much of this gas has accreted relatively pristinely and shock-heated to the virial temperature.  

Many of our general trends match those of \citet{vandevoort12}, who explored radial profiles in a variety of haloes in $z=2$ and $z=0$ OWLS simulation outputs \citep{schaye10}, including stacks of $\approx 10^{12} \msolar$ haloes. Both studies find cool metals in greater abundances at low $z$ than at high $z$ throughout the CGM.  In contrast, however, the OWLS simulations show very similar high-$z$ metal abundances between the phases until $0.5 R_{200}$, where our simulations show a substantial separation of abundance throughout.  It is crucial to note however that \citet{vandevoort12} used smoothed particle metallicities \citep{wiersma09b}, which spreads metals over the SPH kernel and results in a greater mixing of metals between phases.  The main EAGLE simulations \citep{schaye15} also use smoothed metallicities, but our implementation of NEQ ionization and cooling uses discrete, unsmoothed metallicities tied to individual SPH particles, which results in less mixing between phases.

\subsection{Velocity} \label{sec:velocity}

\begin{figure*}%[H]
    \includegraphics[width=0.49\textwidth]{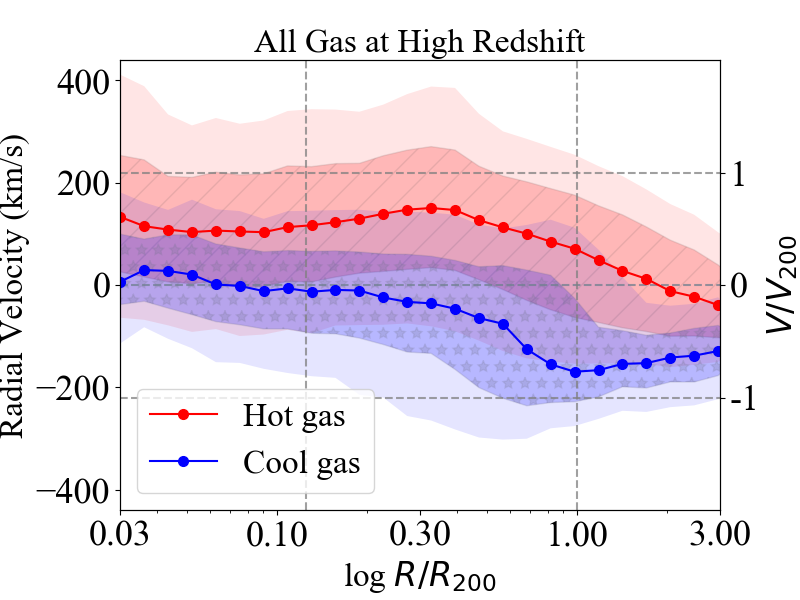}
    \includegraphics[width=0.49\textwidth]{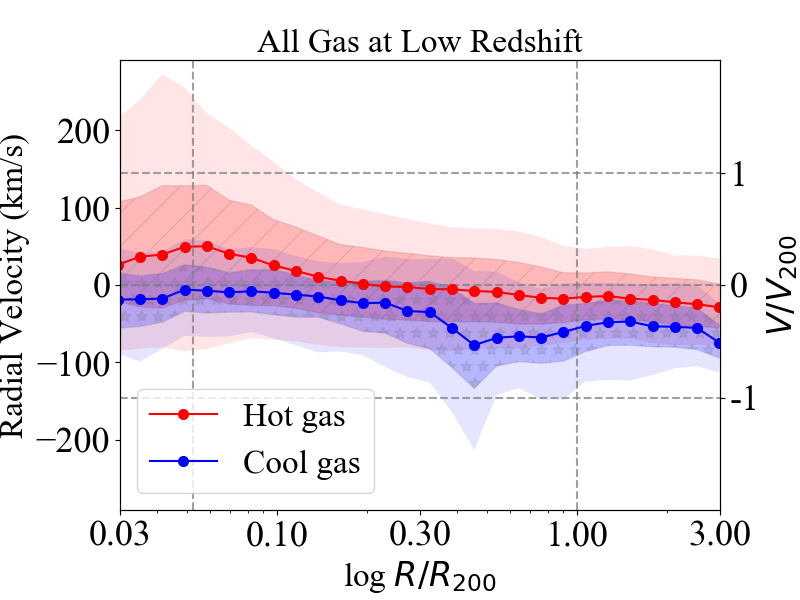}
    \caption{Radial velocity medians (lines) across high-$z$ (right) and low-$z$ (left) halo samples divided into hot ($T \geq 10^{5}$ K; red) and cool ($T < 10^{5}$ K; blue) gas. Inter-quartile (25-75\%) ranges in striped and starred shading, and 10-90\% spreads shown in lighter, solid shading. Horizontal dashed lines represent $v=\pm v_{200}$, and 0 $\kms$ (left axis scale), and vertical dashed lines represent $10$ kpc and $R_{200}$.  Hot gas velocities at high $z$ indicate strong outflows.  High-$z$ cool gas shows net accretion.  Velocities are significantly lower at low $z$, and hot gas transitions from weak outflow to slight inflow at larger radii.}
\label{fig:velradmass}
\end{figure*}

We now discuss velocities, first dividing them into radial and tangential components. The net radial velocity is defined as 
\begin{equation}
    v_{\rm rad} = \frac{{\bm v} \cdot {\bm R}}{R},  
\end{equation}
where ${\bm v}$ and ${\bm R}$ are the velocity and radial position vectors relative to the central galaxy.  Figure \ref{fig:velradmass} plots the medians of cool and hot gas in the high-$z$ and low-$z$ haloes in the left and right panels, respectively.  Dark shading shows the 25-75\% (inter-quartile) range, and light shading shows the 10-90\% range.   At high $z$, most cool gas is inflowing while most hot gas is outflowing.  Horizontal dashed lines show the typical virial velocity $v_{200}\equiv \sqrt{G M_{200}/R_{200}}$, which is $220 \kms$.  Most cool gas inflows slower than $|v_{200}|$, which represents the approximate gravitational speed limit of infalling gas.  Hot gas is typically outflowing at high $z$, with at least 25\% of the gas with $v> v_{200}$ at $R<0.5 R_{200}$.  This indicates that much of the hot gas at high $z$ is associated with superwind outflows driven by the thermal stellar and AGN superwind prescriptions in EAGLE.  It is not clear if hot outflows  above $v_{200}$ escape the halo, but positive net velocities are seen out to $1.5 R_{200}$ at high $z$ in stark contrast to the cool gas, which is dominated by inflows in the outer halo and beyond.  

At low $z$, the velocities are much smaller when considering absolute values and even less so with relative virial values, where the typical $v_{200}=146 \kms$.  There is a net inflow of cool gas, but at lower fractional virial values.  Hot gas shows a net outflow in the interior 30 kpc, but very rarely is there gas moving in excess of $v_{200}$.  

We compare our haloes to those of \citet{vandevoort12}, who plotted cool and hot radial velocities from OWLS \citep{schaye10} simulations that used a kinetic wind prescription for stellar feedback (and no AGN feedback scheme).  At $z=2$, their resulting $v_{\rm rad}$ profile shows similar trends as us at $M_{200}\sim 10^{12} \msolar$.  Hot gas outflows far beyond the virial radius, then reverses to primarily  inflows at $2-3 R_{200}$, the same trend we see.  Their hot gas achieves a maximum velocity in excess of $100 \kms$ at $\sim 0.3 R_{200}$.  However, by low redshift we find higher $v_{\rm rad}$ for hot gas than \citet{vandevoort12}, indicating hot outflows are more prevalent in the EAGLE thermal wind prescriptions for stellar and AGN feedback.  

Cold gas flows inward throughout the entire halo, but reaches a maximum median inflow velocity at $0.8 R_{200}$ at high $z$ and $0.5 R_{200}$ at low $z$.  The same feature is seen in \citet{keres05} (their fig. 19) and \citet{vandevoort12}, and indicates cold accretion decelerating due to weak shocks that do not heat the gas into the hot phase.  The fact that we see the same trends as the simulations without feedback in \citet{keres05} suggests cool accretion operates in a similar fashion despite the presence of feedback.  

\begin{figure*}%[H]
    \includegraphics[width=0.49\textwidth]{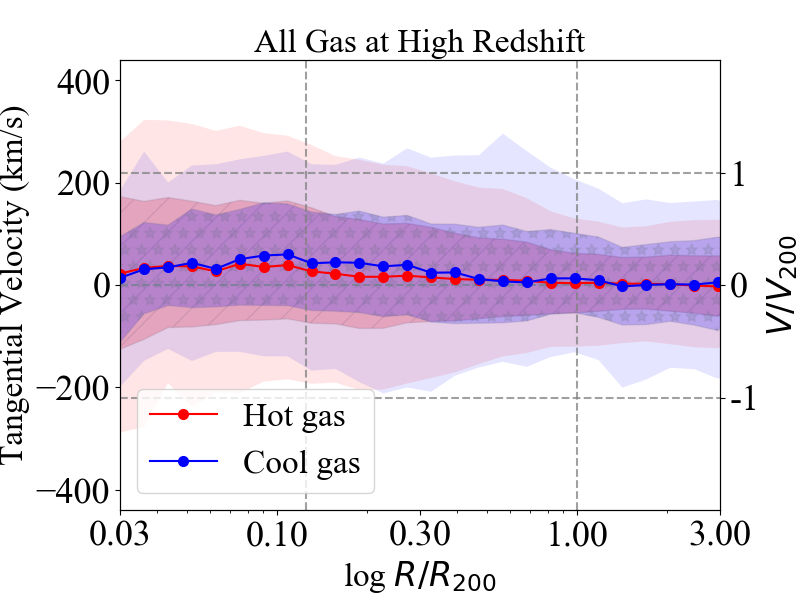}
    \includegraphics[width=0.49\textwidth]{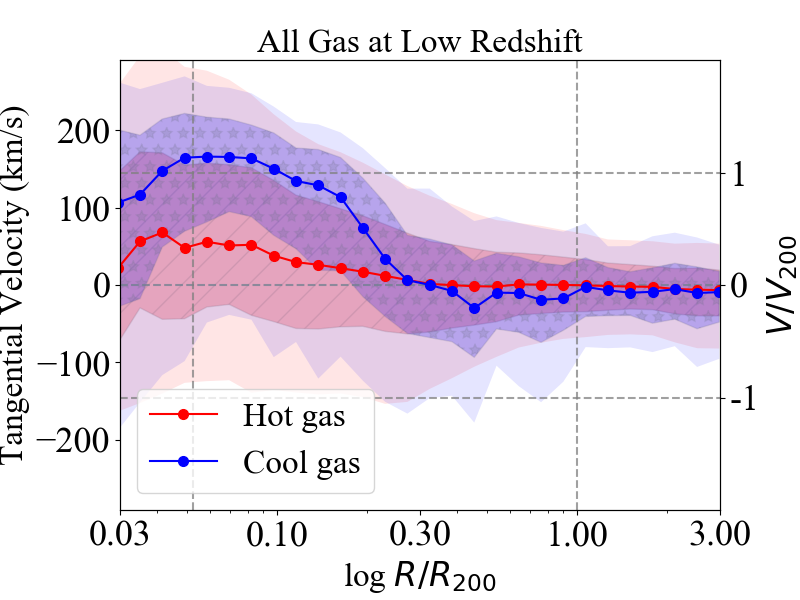}
    \caption{Tangential velocities plotted as in Fig. \ref{fig:velradmass}.  Directional velocities are plotted relative to the stellar angular momentum axis.  Low-$z$ cool ($T < 10^{5}$ K) gas indicates significant rotation in the inner $0.2 R_{200}$, consistent with disc-like structures extending into the CGM.  Low-$z$ hot ($T \geq 10^{5}$ K) gas shows net positive rotation, while at high $z$ there is no strong directionality to the rotation.}
\label{fig:veltanmass}
\end{figure*}

In Figure \ref{fig:veltanmass} we show net tangential velocities, defined as the normalized velocity cross product with radius,  
\begin{equation}
    v_{\rm tan} = \frac{ {\bm v}\times {\bm R}}{R}.
\end{equation}
Note that because we have rotated our haloes to be aligned with the angular momentum vector of the stars within 30 kpc, the sign of the $v_{\rm tan}$ is a measure of correlated/anti-correlated velocities relative to the galaxy's stars if it is positive/negative.  At high $z$ (left panel), there does not appear to be much organization for tangential velocity, which may also stem from the high-$z$ galaxies not having as much organized CGM structure.  The cool gas has slightly greater tangential motion as indicated by the broader dispersion.  

However, by $z=0$ (right panel), tangential velocities show an organized structure of the cool gas indicating a primarily rotationally supported disc extending out to $R \approx 40$ kpc.  Present-day hot haloes also show higher tangential velocities than their high-$z$ counterparts, and the tangential motion inside $0.3 R_{200}$ indicates co-rotation with the galaxy's preferred axis.  \citet{opp18b} showed that these same low-$z$ haloes deviate significantly from hydrostatic equilibrium owing primarily to significant tangential support of the inner hot halo, which in part exhibit sub-centrifugal rotation but also have uncorrelated tangential motions.  In contrast, the high-$z$ haloes do not show evidence for such tangential support in their hot haloes.  The high-$z$ CGM does not show indications of dynamical stability, which contrasts with low-$z$ hot haloes at $r\ga 50$ kpc that are mainly supported by a thermal pressure gradient \citep{opp18b}.  

Tangential velocities were also explored using the Illustris-TNG simulations by \citet{defelippis20}, where they also found higher cool than hot $v_{\rm tan}$ that greatly increase inside $0.5 R_{200}$ for $z=0$ $10^{11.75-12.25} \msolar$ haloes.  They divide their sample into quartiles based on specific stellar angular momentum, $j_*$, and examine all $L^*$ centrals in Illustris-TNG, while excising gas bound to satellites.  Our low-$z$ haloes  preferentially host spiral galaxies, which suggest they have higher $j_*$ than the typical $L^*$ central.  However, our galaxies are unlikely to all be within the highest quartile of $j_*$ in the EAGLE volume, which is the quartile for which \citet{defelippis20} finds the greatest tangential velocities.  

\begin{figure*}%[H]
    \includegraphics[width=0.49\textwidth]{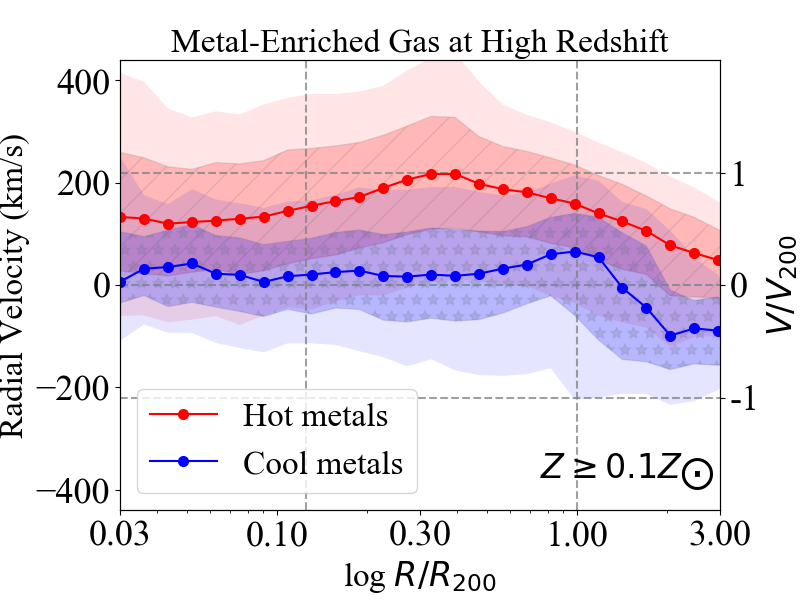}
    \includegraphics[width=0.49\textwidth]{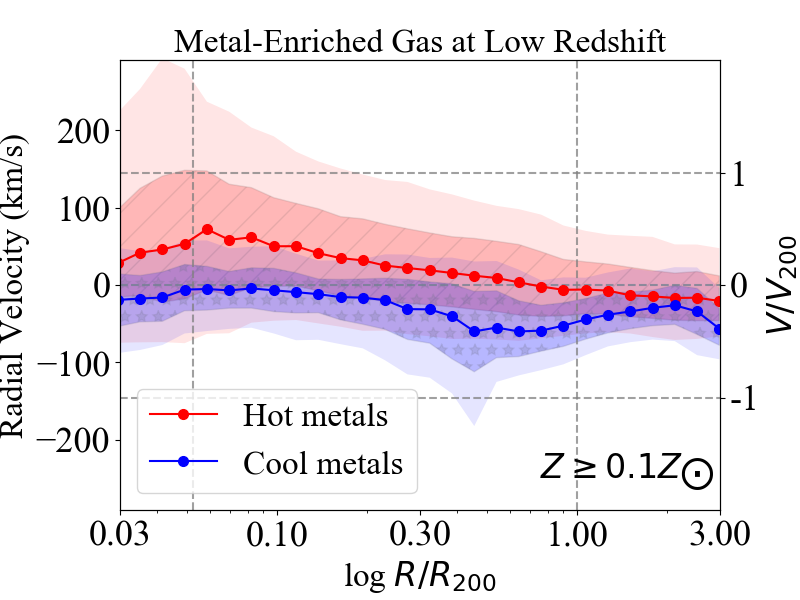}
    \caption{Medians of radial velocities of metal-enriched gas ($Z \geq 0.1 \Zsolar$) as plotted in Fig. \ref{fig:velradmass}. Inter-quartile (25-75\%) ranges in striped and starred shading, and 10-90\% spreads shown in lighter, solid shading. Hot metals at high $z$ are strongly outflowing and show weaker outflows at low $z$.  Cool ($T < 10^{5}$ K) metals are preferentially flowing outward at high $z$ and inward at low $z$.}
    \label{fig:velradmetal}
\end{figure*}

\begin{figure*}%[H]
    \includegraphics[width=0.49\textwidth]{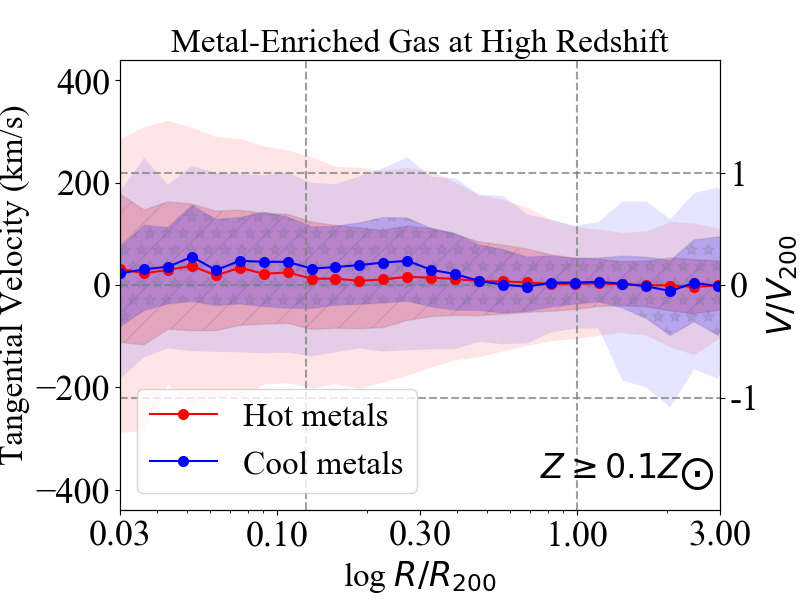}
    \includegraphics[width=0.49\textwidth]{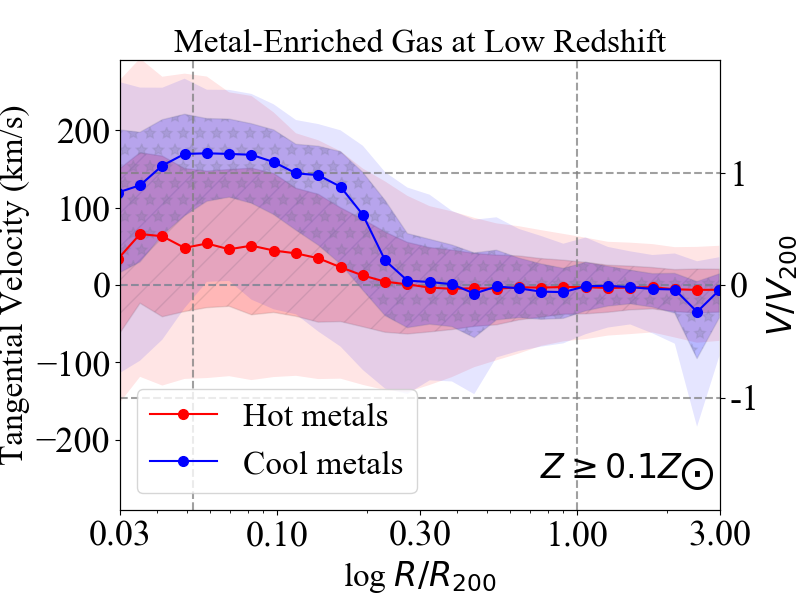}
    \caption{Medians of tangential velocities of metal-enriched gas ($Z \geq 0.1 \Zsolar$) as plotted in Fig. \ref{fig:veltanmass}. Inter-quartile (25-75\%) ranges in striped and starred shading, and 10-90\% spreads shown in lighter, solid shading. Metals show similar medians as gas.  At high $z$, tangential spreads are less for the metals, which are preferentially outflowing. At low $z$, spreads are similar to gas, indicating that metals are well-mixed.}
    \label{fig:veltanmetal}
\end{figure*}

In Figures \ref{fig:velradmetal} and \ref{fig:veltanmetal}, we present the radial and tangential velocities but including only metal-enriched gas, which in this case are gas particles with $\geq 0.1 \Zsolar$ metallicity.  At high $z$, the hot metals show slightly faster moving outflows, reaching a median of 200 $\kms$ near $0.3 R_{200}$, as shown in Fig. \ref{fig:velradmetal} (left panel).  This indicates that metals are preferentially being transported to a large fraction of the halo radius, often becoming ejected from high-$z$ haloes.  The flow of cool metals is also different from that of the total cool gas with median outflows near $0 \kms$ and even positive approaching $R_{200}$. This result contrasts with the cool gas in Fig. \ref{fig:velradmass} that indicates primarily inflowing gas.  We predict UV absorption kinematics of cool gas (e.g. $\HI$) and cool metals (e.g. $\CII$, $\SiIII$, $\MgII$) to be different at high-$z$.

By low redshift, the radial velocity profiles between high-metallicity gas (Fig. \ref{fig:velradmetal}, right panel), and all gas (Fig. \ref{fig:velradmass}, right) exhibit similar shapes and trends.  One difference is that the hot metals are outflowing at higher velocities than the corresponding hot gas, with a net outflow continuing all the way to $0.7 R_{200}$, whereas the hot gas only shows a net outflow to $0.2 R_{200}$.

\citet{turner17} also explores radial velocities at $z\approx 2$ using the main EAGLE volume, finding net inflows for gas, $\HI$, and even metal species ($\CIV$, $\SiIV$) from beyond $1$ Mpc to at least $70$ kpc in their fig. 8.  Our high-$z$ zooms suggest more of a net outflow in the outer CGM where our plot overlaps theirs, but this heavily depends on the cut applied to metals.  If a $Z\geq\Zsolar$ cut is used instead for Fig. \ref{fig:metalradial}, strong radial outflows extend beyond $R_{200}$.  Lower metallicity thresholds result in greater inflows, hence the \citet{turner17} result suggests metal ions arise primarily from lower metallicity gas although there are differences in simulation resolution and halo selection.  

\begin{figure*}
    \includegraphics[width=0.49\textwidth]{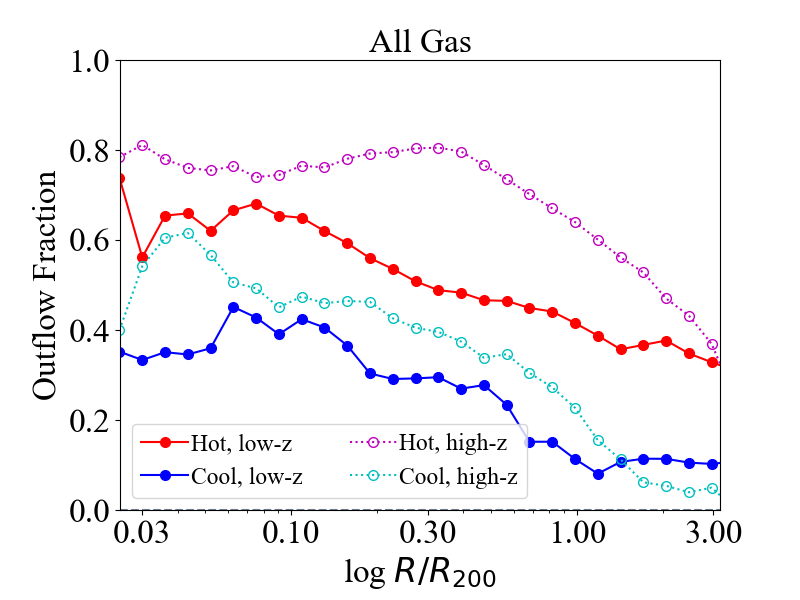}
    \includegraphics[width=0.49\textwidth]{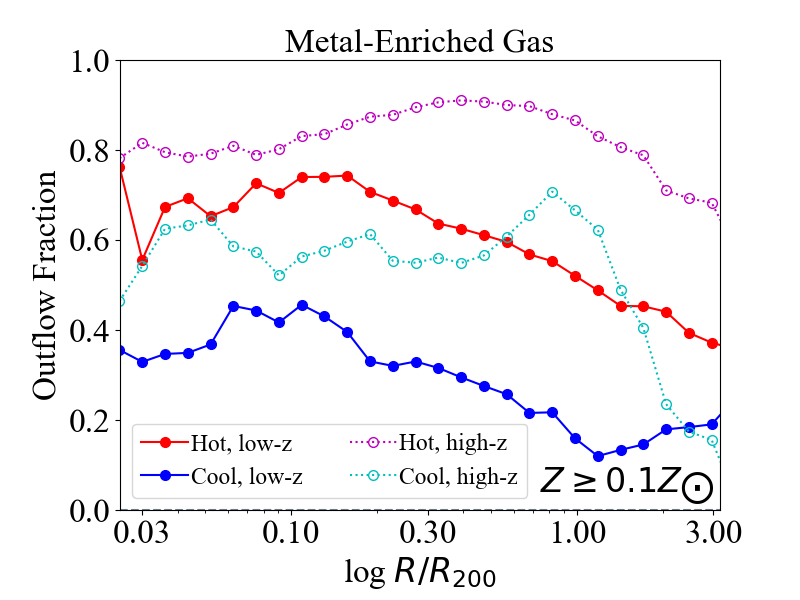}
    \caption{Outflowing fraction of all gas (left) and metal-enriched gas ($Z \geq 0.1 \Zsolar$; right) of all high-$z$ (dotted) and low-$z$ (solid) haloes.  Hot gas ($T \geq 10^{5}$ K) is more dominated by outflows than cool gas ($T < 10^{5}$ K), as is high-metallicity gas relative to all gas.  Cool high-$z$ gas is generally inflowing, but the metals indicate more outflows. }
    \label{fig:outflowfraction}
\end{figure*}

Moving on to tangential velocities of the metals in Figure \ref{fig:veltanmetal}, both cool and hot metals show similar profiles at high $z$ (left panel), but exhibit less overall tangential motion than the high-$z$ gas (cf. Fig. \ref{fig:veltanmass}).  This indicates that metals are on a preferentially radially outflowing trajectory at high $z$.  By low $z$ (right panel), the tangential profiles of cool and hot metals are essentially statistically indistinguishable from gas, indicating that metals are well-mixed throughout the CGM and exhibit specific angular momentum profiles that are similar to gas (see \S\ref{sec:angularmom}). 

Finally, we sum up the results of mass and metals by showing the outflow fraction in Figure \ref{fig:outflowfraction}, where we plot the fraction of gas with net positive radial velocity across our four main subdivisions (high $z$/low $z$ and hot/cool) for all gas (left panel) and high-metallicity gas (right panel). In general, hot gas is more outflowing than cool gas, particularly at high $z$. Cool gas has a similar inflowing proportion at both epochs. Metal-enriched gas has comparatively more outflows in all cases.  The most obvious divergence in trends is that cool, high-$z$ metal-enriched indicates more outflows in contrast with its primarily inflowing nature at low $z$.  Cool gas probed by UV absorption lines may exhibit larger absolute velocities at high $z$ than low $z$, which could be a signature of outflows.   

\citet{hafen19} analyzed FIRE-2 simulations focussing on the origin of the CGM at $z=0.25$ and $z=2$ via particle tracking of individual gas elements.  Their $\sim 10^{12} \msolar$ haloes exhibit many of the trends we see here, including winds from the central galaxy extending much further into the CGM at $z=2$ than at $z=0.25$.  Their use of tracking finds that much more of the $z=2$ CGM gas originates from central galaxy winds than at $z=0.25$, where the dominant origin of CGM gas is accretion from the IGM (their fig. 9).  Like our haloes, their $\sim 10^{12} \msolar$ haloes at low $z$ are dominated by hot gas while their highest mass $z=2$ haloes ($\sim 10^{11.7} \msolar$) show more of an equitable split between cool and hot phases (their fig. A1).  Our lower velocities at low-$z$ indicate that gas cycles through the CGM significantly more slowly than at high-$z$, which agrees with the \citet{hafen20} finding that half of  the FIRE-2 low-$z$ CGM remains within the virial radius as CGM gas for $\sim 3$ Gyr, while most of the $z=2$ CGM gas will either accrete onto the galaxy or be ejected from the halo within a Gyr (their figs. 2 and 5).  The vast majority of their $z=0.25$ CGM gas that remains in the CGM is hot, while the cool CGM more likely accretes onto the central or a satellite (their fig. 6) and \citet{hafen19} finds the cool gas is more aligned along the disc of the galaxy as opposed to a more spherical distribution of the hot gas.  Our haloes retain fewer baryons overall inside the virial radius than FIRE-2, which have higher low-$z$ stellar fractions \citep[][their fig. 1]{hafen19} suggesting that our CGM gas is less likely to be accreted onto the central galaxy and more likely to be ejected from the CGM.  This in part owes to the presence of AGN feedback in our simulations, which is absent in the FIRE-2 simulations.

\subsection{Angular Momentum} \label{sec:angularmom}

\begin{figure*}%[H]
    \includegraphics[width=0.49\textwidth]{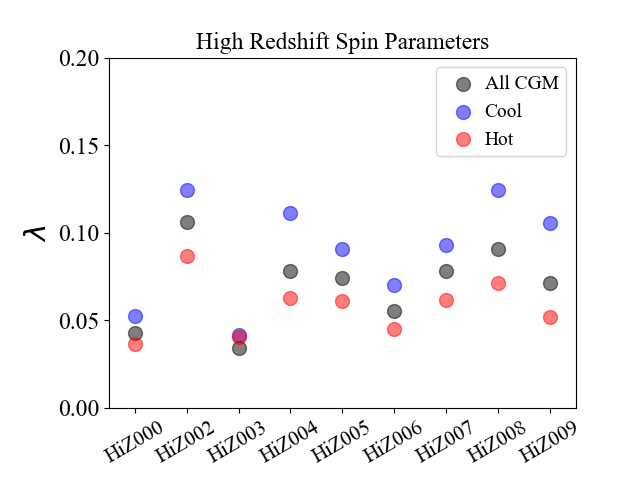}
    \includegraphics[width=0.49\textwidth]{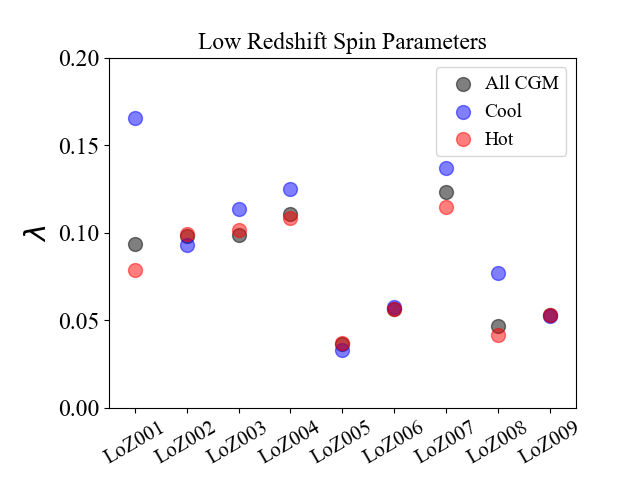}
    \caption{Angular momentum spin parameters of the 9 high-$z$ (left) and 9 low-$z$ haloes (right).  All CGM gas is plotted as black points, cool ($T < 10^{5}$ K) gas as blue points, and hot ($T \geq 10^{5}$ K) gas as red points.  Hot gas always has lower spin parameters than cool gas at high $z$, and spin parameters have higher averages at low $z$.  Given that most of the CGM mass is hot at low $z$, there is more angular momentum in the hot phase than the cool phase.}% Note that LoZ006 and LoZ009 have identical overlapping spin parameters.}
    \label{fig:spinparams}
\end{figure*}

The angular momentum of the CGM has significant implications for the gas that accretes onto a galaxy, forms stars, and builds a galaxy's morphology.  In our selection of star-forming galaxies at high and low $z$, we derive halo spin parameters,  
\begin{equation}
\lambda = \frac{j}{ \sqrt[]{2} R_{200} v_{200}},
\end{equation}
where the specific angular momentum $j$ is defined as
\begin{equation}
j = \frac{\| {\bm J} \|}{\sum\limits_{i} m_{i}},
\end{equation}
and ${\bm J}$ is the angular momentum vector sum,
\begin{equation}
{\bm J} = \sum_i m_{i} {\bm v}_{i}\times {\bm R}_{i}.
\end{equation}
\noindent over particle indices $i$.  In Figure \ref{fig:spinparams}, we plot the halo spin parameters for both high and low $z$.  In both samples, the cool phase of the CGM tends to have more angular momentum than the hot phase.  The median total CGM spin parameters and interquartile spreads are $\lambda=0.074^{+0.004}_{-0.019}$ ($0.094^{+0.005}_{-0.041}$) at high (low) $z$.  A notable difference is the decline in the spread of angular momenta between cool and hot gas at later times.  At high $z$, the median $\lambda_{\rm hot}$ is $0.061$, which is $65\%$ of the median $\lambda_{\rm cool}=0.093$.  By low $z$, the difference is less with $\lambda_{\rm hot}=0.079$, which is 85\% of $\lambda_{\rm cool}=0.093$.  The $\lambda_{\rm cool}$ values are well within the range of previous studies that show $\lambda_{\rm cool}$ to be several times that of the dark matter, $\lambda_{\rm DM}$ \citep{stewart11,stewart17}.  \citet{stevens17} showed that EAGLE haloes in general have higher $\lambda_{\rm hot}$ than $\lambda_{\rm DM}$, and \citet{opp18b} showed that these low-$z$ haloes had $\lambda_{\rm hot} = 3\times \lambda_{\rm DM}$.   

The low-$z$ combination of the total mass of the CGM being dominated by the hot phase and $\lambda_{\rm hot}$ being a high fraction of $\lambda_{\rm cool}$, means that the median angular momentum of the hot halo out to $R_{200}$ is about $5\times$ that of the cool CGM, with values of ${\bm J}_{\rm hot}=2.4\times 10^{14}$ and ${\bm J}_{\rm cool}=4.7\times 10^{13}\ \Msolkmskpc$ respectively.  The low-$z$ hot CGM is the largest repository of angular momentum of any phase.  Most of the hot angular momentum is spatially extended with $<10\%$ of the ${\bm J}_{\rm hot}$ coming from $<0.3 R_{200}$, which contrasts with ${\bm J}_{\rm cool}$ for which the proportion from $<0.3 R_{200}$ is half.  At high $z$, the cool CGM has total angular momentum 75\% higher than the hot CGM (cf. ${\bm J}_{\rm cool}=8.1\times 10^{13}$ vs. ${\bm J}_{\rm hot}=6.0\times 10^{13}\ \Msolkmskpc$), in large part due to the hot CGM being primarily outflowing and not rotating.  If the ISM criterion uses only SFR$>0$, instead of our definition described in \S\ref{sec:phasedef}, then ${\bm J}_{\rm cool}$ becomes $1.0\times 10^{14} \ \Msolkmskpc$ at high-$z$ but negligible difference at low-$z$.  

\begin{figure*}%[H]
    \includegraphics[width=0.49\textwidth]{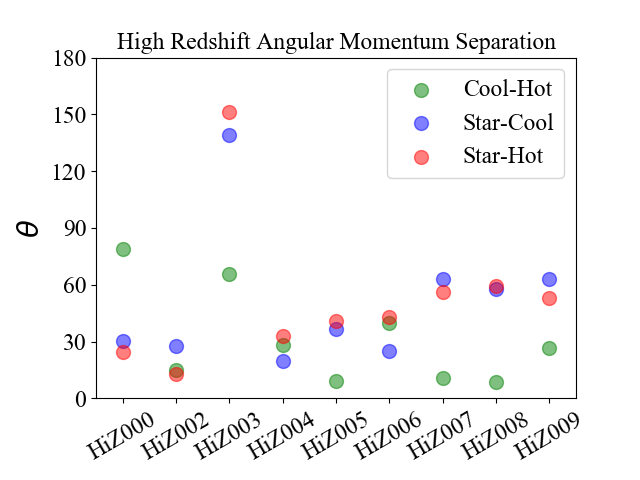} 
    \includegraphics[width=0.49\textwidth]{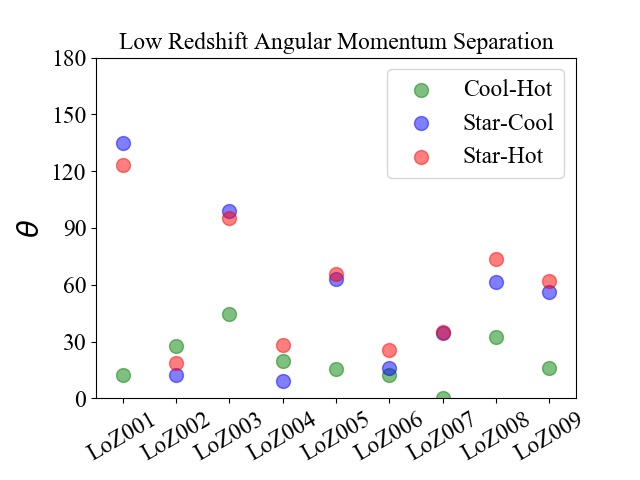}
\caption{The angles of separation between the hot ($T \geq 10^{5}$ K) and cold ($T < 10^{5}$ K) gas angular momentum vectors (Cool-Hot) are plotted as green points.  The angle between the cool (hot) CGM angular momentum vectors and the stellar angular momentum vectors are plotted as blue (red) points.  Haloes at high and low $z$ exhibit well-aligned angular momentum for their hot and cool CGM, but significant mis-alignment between the angular momenta of the CGM and stellar disc is typical at both high and low $z$.  Low-$z$ galaxies with all vectors well-aligned often have grand design spiral morphologies, including halo LoZ004 shown in the right panel of Fig. \ref{fig:maps}.}
\label{fig:spinangles}
\end{figure*}

We also consider angular separation $\theta$ in degrees between the angular momentum vectors of the cool CGM, hot CGM, and the stellar component in Figure \ref{fig:spinangles}. The cool and hot CGM are fairly well-aligned with a median angle of $\theta_{\rm cool-hot}=27^\circ$ ($16^\circ$) at high (low) $z$. However, the alignment is less between the stellar component (all stars within 30 pkpc) and the CGM with median $\theta_{\rm star-cool}=37^\circ$ ($56^{\circ}$) and $\theta_{\rm star-hot}=43^\circ$ ($62^\circ$).

%The cool and hot CGM are well-aligned with a median angle of $\theta_{\rm cool-hot}=18^\circ$ at both high $z$ and low $z$.

\citet[][their Fig. 14]{stevens17} also looked at angles between cool and hot gas, finding a somewhat greater offset between the cool and hot CGM than our haloes, though this may be in part due to their differing definitions of their ``cold'' gas, where they include ISM, plus our selection of only star-forming galaxies.  Nevertheless, there is strong alignment between the cool and hot CGM, and more randomness with the orientation of the stellar disc, which represents the integrated result of accretion and star-formation over all previous epochs.  We check that the greater star-cool angles do not contradict the cool CGM rotating disc-like structure aligned with the stellar component at low $z$ in Fig. \ref{fig:veltanmass} that show strong co-rotation between the stellar and inner ($\la 0.2 R_{200}$) cool CGM.  Mis-alignment often arises from extended cool CGM structures with significant angular momentum at large radii and no correlation with the stellar disc.  On the other hand, some of the most well-aligned galaxies along all three vectors in Fig. \ref{fig:spinangles} (e.g. LoZ002, LoZ004, LoZ007) appear as grand design spirals with extended CGM discs (cf. Fig. \ref{fig:maps}, right panels).  

\citet{defelippis20} finds stronger alignment between the stellar and CGM angular momenta in their highest $j_*$ quartile with $\theta_{\rm star-cool}\approx 15^\circ$ and $\theta_{\rm star-hot}\approx 27^\circ$ in Illustris-TNG (their fig. 2).  Their lowest $j_*$ quartile shows angles of $\approx 60^\circ$, more similar to our results.  The existence of extended CGM structures, including the CGM associated with satellites that is excised by \citet{defelippis20}, likely biases high our angles between stars and the CGM.  Like us, they find that the hot specific angular momentum, $j_{\rm hot}$, is a significant fraction of the $j_{\rm cool}$ at both $z=0$ and $z=2$.  

\begin{figure*}
\includegraphics[width=.495\textwidth]{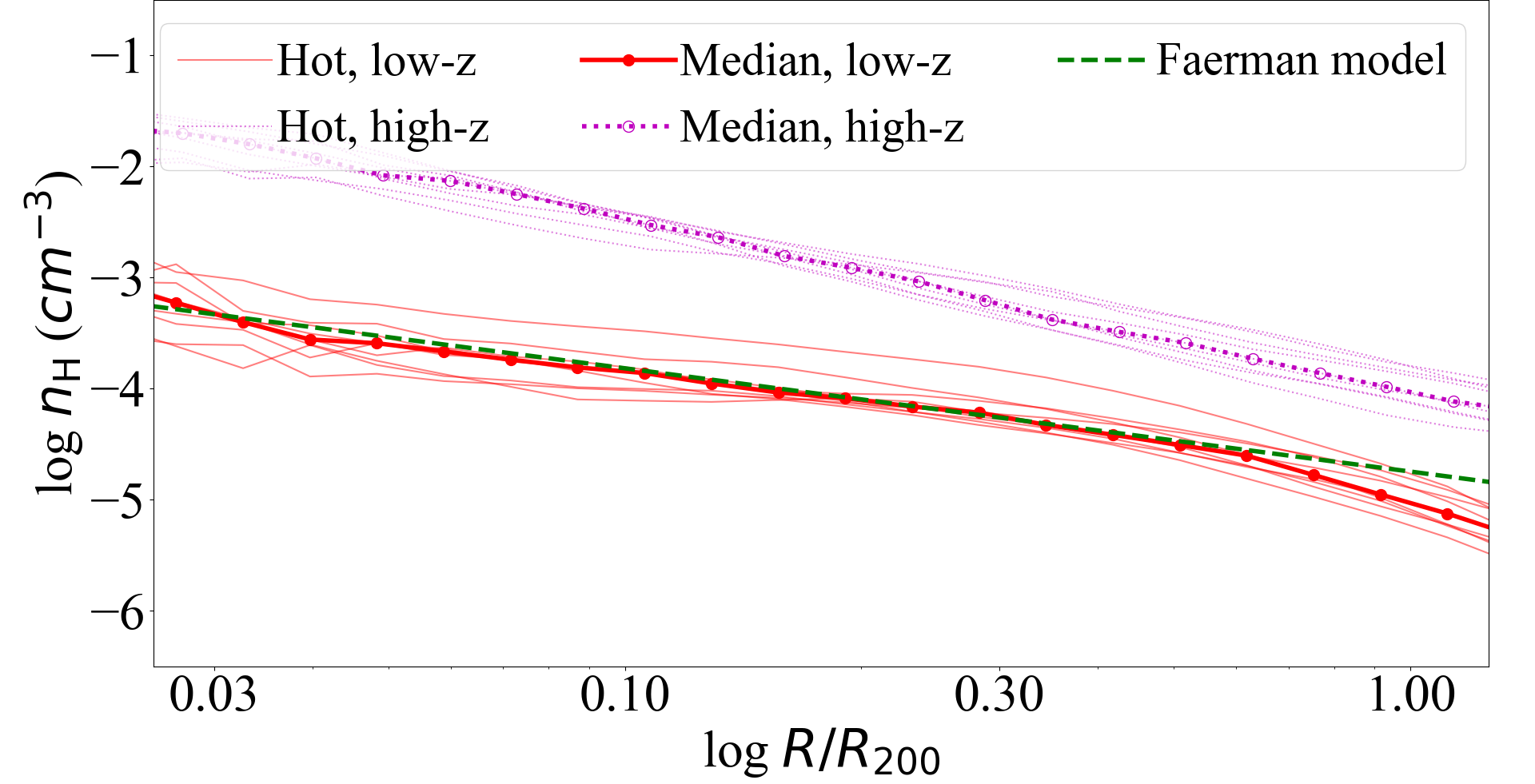}
\includegraphics[width=.495\textwidth]{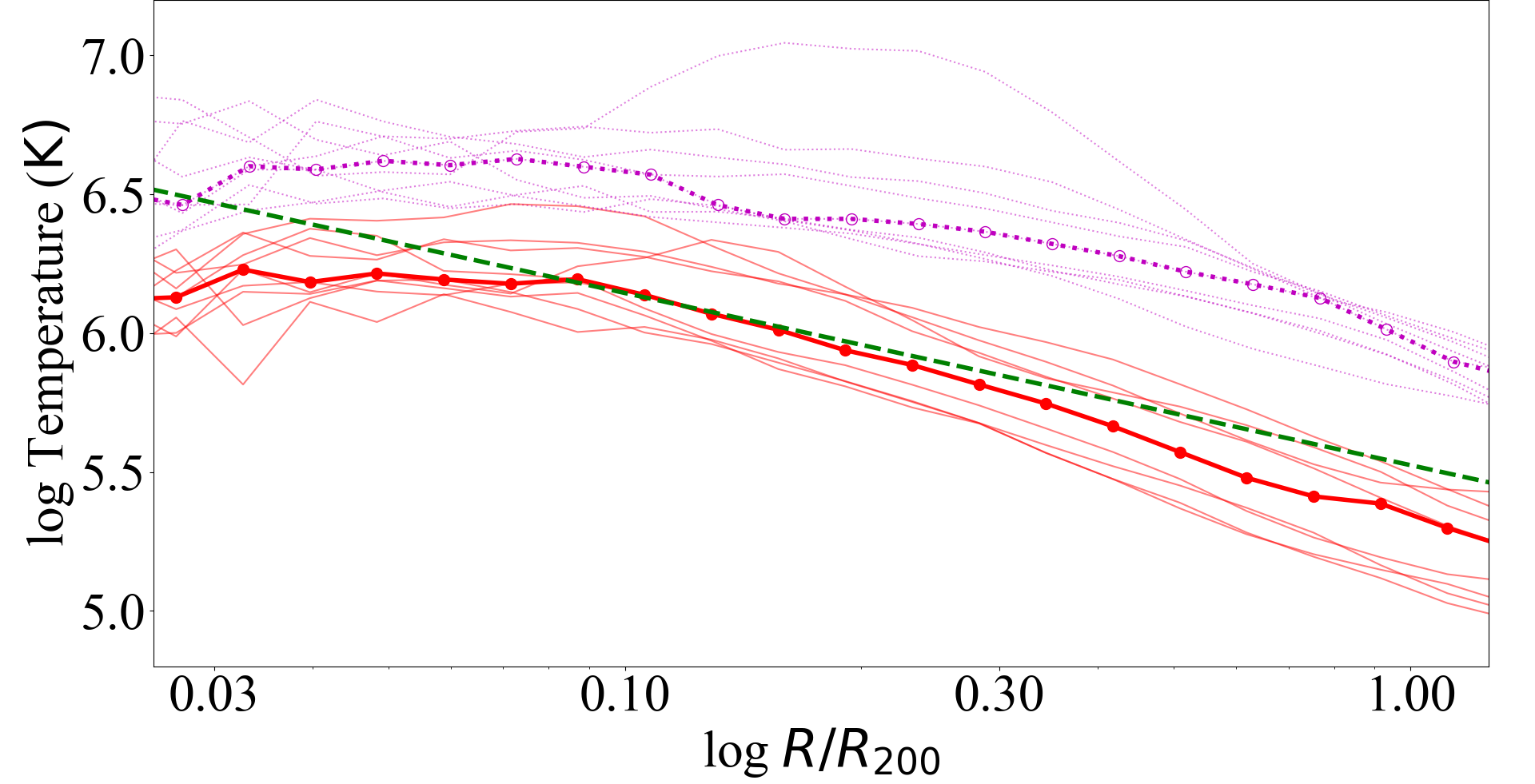}
\includegraphics[width=.495\textwidth]{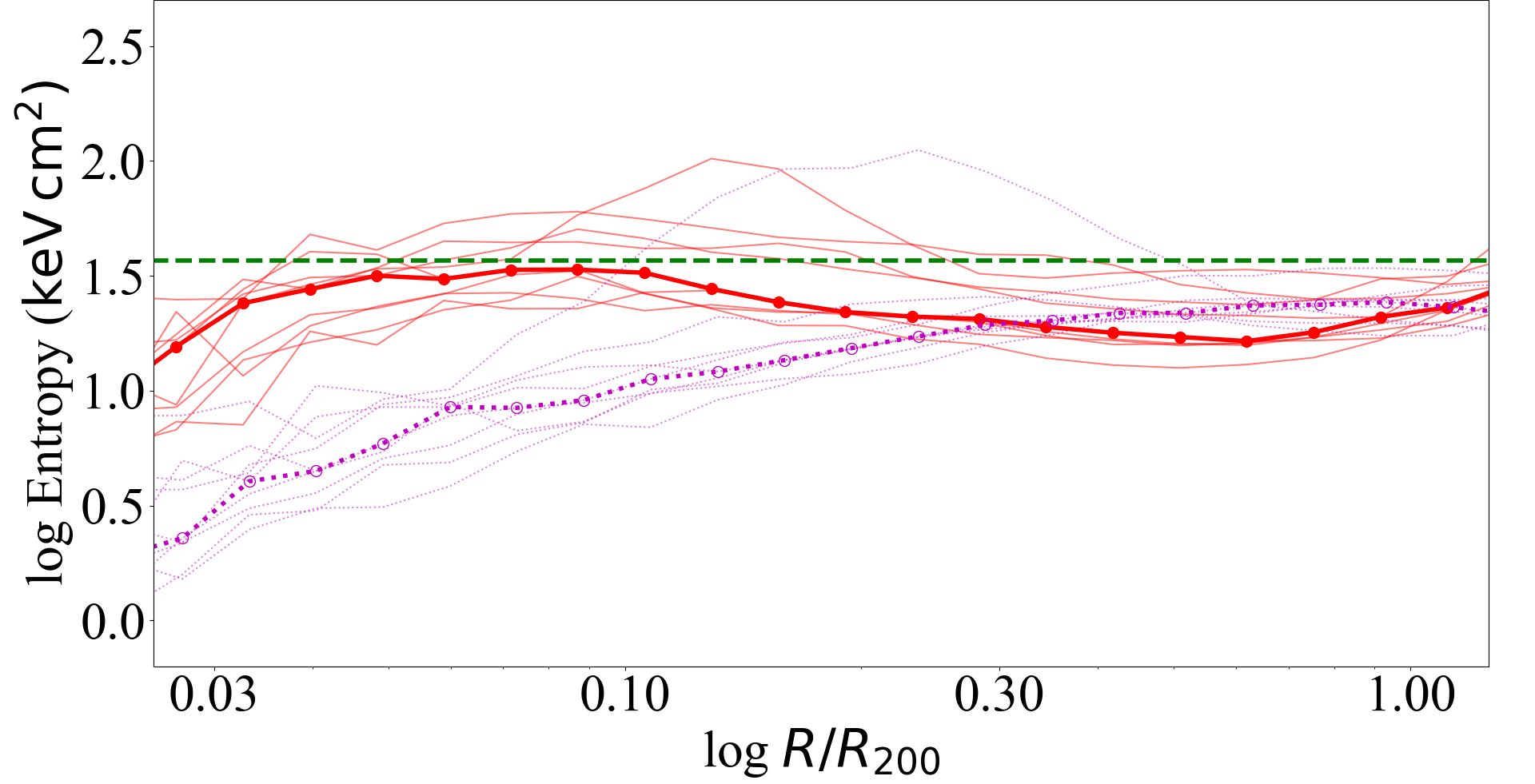}
\includegraphics[width=.495\textwidth]{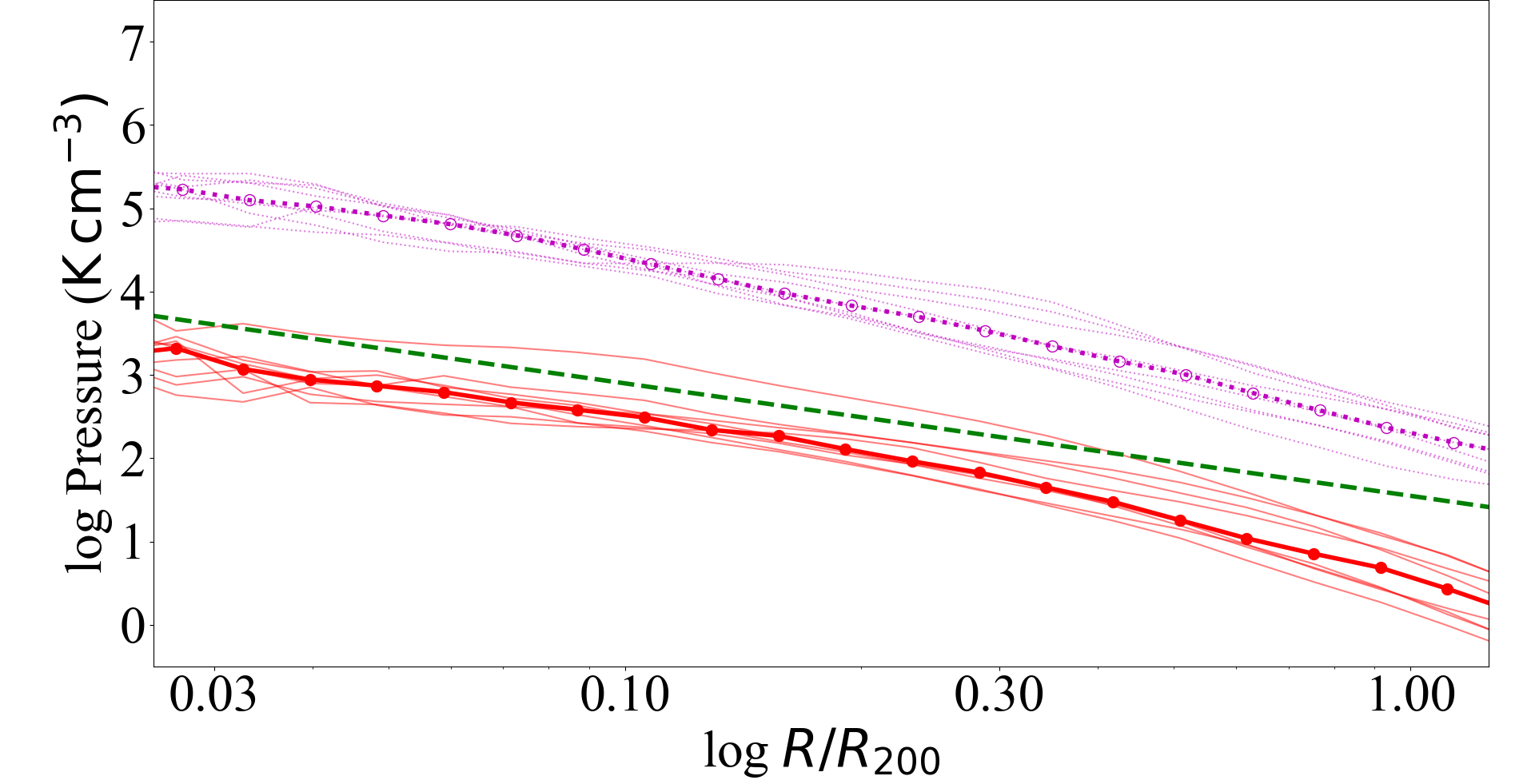}
\caption{Volume-weighted radial trends of the density, temperature, entropy, and pressure for hot phase gas at high $z$ (dotted magenta) and low $z$ (solid red). Stacked medians of the profiles are shown with thicker lines. The \citet{faerman20} isentropic model is shown in dashed green for a $z=0$ Milky Way-like hot halo.
}
\label{fig:hotprofiles}
\end{figure*}

\subsection{Hot gas radial profiles} \label{sec:hotgas}

The hot gas dominates the CGM volume around $L^*$ galaxies \citep[e.g.][]{bregman07,stocke13}, therefore we plot volume-weighted radial profiles of this component in Figure \ref{fig:hotprofiles}.  The gas density (upper left panel) shows flatter radial profiles at low $z$ than at high $z$ with the following fits to the medians of the 9 haloes between $R=0.1$ and $1.0 R_{200}$, using 
\begin{equation}
    \nh = n_{{\rm H}, R_{200}} \left(\frac{R}{R_{200}}\right)^{-\alpha_{\nh}}
\end{equation}
where $\alpha_{\nh} = 1.6$ at high $z$ and becomes $1.1$ at low $z$.  $n_{{\rm H}, R_{200}}=9.0\times 10^{-5}$ and $1.4\times 10^{-5} \cmc$ at the two respective redshifts.  The temperature profiles flatten towards the center, and decline with power laws fit between $0.1-1.0 R_{200}$ using 
\begin{equation}
    T = T_{R_{200}} \left(\frac{R}{R_{200}}\right)^{-\alpha_T}.  
\end{equation}
At high (low) $z$, $T_{R_{200}}=1.1\times10^6$ ($2.2\times 10^5$) K and $\alpha_T= 0.5$ ($0.8$).  

%DENSITY MASS-WEIGHTED- where $\alpha_{\nh} = 1.7$ at high $z$ and becomes $1.2$ at low $z$.  $n_{{\rm H}, R_{200}}=3.2\times 10^{-4}$ and $2.8\times 10^{-5} \cmc$ at the two respective redshifts. The temperature profiles flatten towards the center, and decline with power laws fit between $0.1-1.0 R_{200}$ using 
%TEMP MASS-WEIGHTED- At high (low) $z$, $T_{R_{200}}=2.0\times10^6$ ($4.0\times 10^5$ K) and $\alpha_T= 0.5$ ($0.8$).  

We also plot the entropy profiles
\begin{equation}
K = T n_{e}^{-2/3}
\end{equation}
in the lower left panel of Fig. \ref{fig:hotprofiles}, where $n_e$ is the free electron density.  At high $z$, entropy is rising with $R$, and slightly flattening at large $R$.  This gas is strongly outflowing (Fig. \ref{fig:velradmass}), which may indicate high-entropy gas is preferentially traversing outward resulting in this profile.  Entropy-driven winds \citep{bower17, keller20} appear to play an essential role in ejecting gas from high-$z$ haloes that exhibit higher outflow rates through the virial radius than their low-$z$  counterparts in EAGLE \citep[][their fig. 1]{mitchell20}.  At low $z$, entropy rises only in the inner $0.1 R_{200}$, where it also coincides with positive radial outflows in Fig. \ref{fig:velradmass} that are much weaker than at high $z$.  Beyond, the entropy actually falls slightly before it recovers in the outer halo.  This coincides with median hot $v_{\rm rad} \approx 0\  \kms$, which indicates a different source of the outer hot halo.  Much of this gas has very low metallicity (Fig. \ref{fig:metallicity}), which suggests a source of accretion from the IGM.  It is curious that rising entropy profiles often coincide with outflowing gas at both both epochs, because such entropy profiles are also indicative of dynamically stable configurations in the centers of clusters, although such profiles are typically steeper \citep[e.g.][]{voit05}.     %In sum, the low-$z$ outer halo has a large range of metallicities along with a spread in $v_{\rm rad}$, thus indicating contributions from both inflows and outflow.  

Finally, we plot pressure profiles,
\begin{equation}
P= n_{\rm H} T,
\end{equation}
in the lower right panel, where high-$z$ pressures are higher everywhere than at low $z$.  The pressure at fixed $M_{200}$ should scale approximately as $R_{200}^{-4}$, which gives the $\approx 30-40$ factor in pressure difference between low and high $z$.  This appears to be the case at $R=0.2 R_{200}$, but the difference is greater at lower (higher) radii where high-$z$ haloes are denser (hotter) than self-similar scaling relations.  

We overplot the green dashed lines in Fig. \ref{fig:hotprofiles} of the \citet{faerman20} $z=0$ isentropic hot halo profiles in all panels.  Their model, developed for a Milky Way-like halo, is a reasonable representation of our flat low-$z$ entropy profile between $0.1-1.0 R_{200}$.  Our densities and temperatures also show good agreement, especially in the inner halo.  Our pressure profile is lower than the \citet{faerman20} model that also includes non-thermal sources of turbulent and magnetic/cosmic ray pressure.  That model assumes hydrostatic equilibrium (HSE), but \citet{opp18b} showed that these low-$z$ haloes are not well-described by HSE in their inner CGMs, though at $\ga 0.5 R_{200}$ the thermal pressure gradient accounts for $\geq 75\%$ of the support against gravity (their fig. 2).  

We also contrast to the \citet{stern19} hot gas steady-state cooling flow models for Milky Way-mass haloes, which have more steeply rising entropy profiles as a result of higher $\alpha_{\nh}$ and lower $\alpha_{T}$.  Resolving X-ray profiles around Milky Way-like galaxies \citep[e.g.][]{li13} as has been done for more massive spirals \citep[][]{anderson16,bogdan17,li17,das19} can help distinguish these contrasting models.  Central X-ray emission from individual galaxies may be detectable with the {\it Chandra} X-ray telescope according to Illustris-TNG simulations \citep{truong20} and the {\it eROSITA} mission should be able to observe extended emission in stacks of haloes as predicted by both the EAGLE and Illustris-TNG simulations \citep{opp20b}.  

\section{Discussion} \label{sec:discussion}  

\subsection{CGM mass contents}

In Table \ref{tab:masses}, we list $f_{\rm CGM}\equiv M_{\rm CGM}/M_{200}(\Omega_{\rm M}/\Omega_{\rm b})$, the total mass content of the CGM (here defined as inside $R_{200}$) normalized to the cosmic baryon fraction.  $f_{\rm CGM}$ averages $0.34$ at high $z$ and is $0.47$ at low $z$.  At high $z$, \citet{pezzulli19} calculated that $f_{\rm CGM}\geq 0.70$ across cool and hot phases are necessary to reconcile giant $\lya$ emission nebulae observed around quasar hosts, which they assume live in haloes of $M_{200}\sim 10^{12} \msolar$.  However, this higher value than our $f_{\rm CGM}$ can be rectified if these quasar hosts have higher halo masses, which both lowers the $f_{\rm CGM}$ that \citet{pezzulli19} calculate in their analysis and raises the $f_{\rm CGM}$ using higher halo masses in EAGLE that generally have higher $f_{\rm CGM}$ \citep{davies19}.  

The low-$z$ average $f_{\rm CGM}$ value is significantly higher than the typical value observed at similar halo mass in the EAGLE Ref volume with $8\times$ lower mass resolution, where $f_{\rm CGM}=0.2$ \citep{davies20}.  We expect higher $f_{\rm CGM}$ values, given that our haloes host star-forming galaxies and \citet{davies19} and \citet{opp20a} showed that these haloes have higher baryon fractions, owing to stellar and black hole feedback preferentially ejecting CGM gas from haloes hosting passive galaxies \citep[see also][for discussion of this effect in the Illustris-TNG simulation]{terrazas20, davies20}.  \citet{opp20a} found $f_{\rm CGM}$ averaged $0.35$ in the highest quartile of sSFR for $M_{200}=10^{12.0-12.3} \msolar$ EAGLE haloes.  This indicates that $f_{\rm CGM}$ is higher at the {\it M5.3} resolution, used also in the Recal-L025N0752, than for the main EAGLE $100^3$ Mpc$^3$ volume.  

\subsection{CGM metal contents}

\citet{peeples14} calculated the expected metal content yielded from stars over cosmic history, finding that the stellar and ISM contributions fell far short of the expected metal content, by at minimum a factor of $2$.  Our "LoZ" simulations confronted this short-fall in O16, arguing that most metals are ejected into the CGM and IGM, often beyond $R_{200}$.  Our average CGM metal contents at $z=0$ are $1.6\times 10^8 \msolar$ for cool metals and $1.2\times 10^8 \msolar$ for hot metals, which are both more than the ISM metal content, $1.1\times 10^8 \msolar$.  The content of metals recycled into later generations of stars is $3.3\times 10^8 \msolar$.  O16 quantified the oxygen content ejected beyond $R_{200}$ at $\approx 35-40\%$ of the expected oxygen yield for $10^{12} \msolar$ haloes, hence we expect $3-5\times 10^8 \msolar$ more diffuse metals beyond $R_{200}$ given the yields and nucleosynthetic sources of metals used in EAGLE.

%However, this does not include metals recycled in later generations of stars nor metals ejected beyond $R_{200}$, which O16 quantified.  The average mass of stars is $1.6\times 10^{10} \msolar$ across our 9 haloes, and the expected yield of metals is $\approx 5\%$, our zoom haloes yield $\approx 8\times 10^{8} \msolar$, of which about half is accounted for in the ISM plus CGM (at $\leq R_{200}$, cf. with fig. 9 in O16).  The rest of the unaccounted metals lie beyond $R_{200}$.  [XXX-JOOP COMMENT OF METAL MASS IN STARS]

\citet{hafen19} found a much greater fraction of the metal yield ends up in stars at $z=0.25$ in FIRE-2 (70-90\%, their fig. 3) than in our haloes (25-35\% at $z=0.2$, O16, their fig. 9).  \citet[][their \S5.1]{opp18c} discussed that EAGLE-CGM simulations have yields that are consistent with \citet{peeples14} but are higher than the ones used in FIRE-1 \citep{muratov17}, where stellar metallicites are similar to O16 but CGM metallicities are much lower.  \citet{hafen19} discussed in their \S4.1.2 that FIRE-2 has similar metal yields as FIRE-1, which are about half as much as used by \citet{peeples14}.  Our simulations yield more metals and place proportionally  more of those metals in diffuse gas, resulting in higher CGM metallicities (Fig. \ref{fig:metallicity}) than FIRE-2 \citep[][their fig. 18]{hafen19}.  \citet{opp18c} found that their higher metallicities are necessary to reproduce COS-Halos low-ion metal absorber statistics \citep{werk13}, but given the uncertainty in ionization corrections it is very possible that fewer metals are necessary to reproduce observed low-ion metal absorbers.  

At high $z$, we find an average of $4.2\times 10^7$, $6.9\times 10^7$, and $2.0\times 10^8 \msolar$ of metals in the cool CGM, hot CGM, and ISM respectively.  This totals to $3.1\times 10^8 \msolar$, which is similar to the amount of metals in stars, $2.8\times 10^8 \msolar$, in these high-$z$ haloes.  A smaller fraction of metals is ejected beyond $R_{200}$, though we save a complete accounting of high-$z$ metals in the context of the missing metals problem at $z=2-3$ \citep[e.g.][]{bouche07} for further work.

%    Metals               Cool        Hot         ISM
%    HiZ Grand average:	    4.33E+07	7.00E+07	1.94E+08
%    LowZ Grand average:	1.62E+08	1.20E+08	1.06E+08    

\subsection{Are low redshift hot haloes rotating?}

We show in Fig. \ref{fig:veltanmass} that low-$z$ hot haloes show a net rotation that is aligned with the stellar axis, but substantially sub-centrifugal as the median $v_{\rm tan}\approx 40 \kms$ inside $0.1 R_{200}$.  \citet{hodges16} calculated a rotation speed of $183\pm 41 \kms$ for the Milky Way's hot halo by measuring $\OVII$ absorption line centroids.  While this is 75\% of the solar rotational speed around the Galactic center of $240 \kms$ \citep{reid14}, our low-$z$ simulations do not exhibit as high values for hot halo co-rotation as the Milky Way, and our hot halo rotational axes are often substantially mis-aligned with the stellar disc (Fig. \ref{fig:spinangles}).  Furthermore, a closer examination of the measured velocities as a function of Galactic latitude and longitude in fig. 5. of \citet{miller16} shows significant scatter and deviations from aligned co-rotation, which they argue requires a much higher resolution X-ray spectrometer to accurately observe.  

\citet{opp18b} found that our low-$z$ inner haloes have significant uncorrelated tangential motions that do not add to the summation of net directional rotation or angular momentum.  Nevertheless, this paper did argue that there existed as much angular momentum in the inner hot haloes as calculated by \citet{hodges16} out to 90 kpc.  The low-$z$ galaxies with larger grand design spiral appearances (LoZ002, LoZ004, LoZ007) have among the highest hot halo spin parameters (Fig. \ref{fig:spinparams}) and aligned spin axes (Fig. \ref{fig:spinangles}).  In the future, it would be useful to consider how total angular momentum in the hot CGM relates to galactic morphology.

%Potential questions to consider here: Why do high redshift haloes have a smaller separation in mass fraction? Why is there more inflowing hot gas than cool gas at all redshifts? Why do cooler metals move faster tangentially, particularly at low redshift?

%At what halo mass is the transition between systems dominated by cold versus hot gas? Is the metallicity of cold gas smaller or larger? and is the spread larger? Is there more outflowing gas in high or low redshift galaxies?

%Most outflowing CGM gas is hot phase, as shown in both redshift regimes in Figure 10. This validates the results of \citep{vandevoort12}, though they expect this not to be the case with higher-mass haloes. The outflowing percentage of cool gas is nearly identical between high and low redshift haloes. Metal flows show a similar trend, with the clear exception of cool phase gas at high redshift which nearly doubles its outflow to 80\%. 

%The cool, low-$z$ CGM masses can be compared to a summation of UV-traced CGM gas compiled at low-$z$ by \citet{werk14}, who estimate $M_{{\rm cool, low-}z}=2\times10^{10} \msolar$, and is updated to be $\sim 10^{11} \msolar$ by \citet{prochaska17}.  The hot CGM masses, compiled by \citet{bregman18} are estimated to be $M_{{\rm hot, high-}z}$.  

\section{Summary}  \label{sec:summary}

We have presented the physical characteristics of circumgalactic haloes simulated at high ($z\approx 2-3$) and low ($z = 0$) redshifts using a set of EAGLE zoom simulations of $10^{12} \msolar$ haloes hosting star-forming galaxies.  These simulations demonstrate the changes in the CGM around $L^*$ galaxies at two epochs separated by 10 Gyr. The mean $M_{200}$ of our 9 high-z (low-z) haloes is $10^{12.04}$ ($10^{12.07}$) $\msolar$. The primary results are as follows:  

\begin{enumerate}{}{\leftmargin=1em}
    \item High-$z$ gaseous haloes have nearly as much cool ($T < 10^{5}$ K) gas as hot ($T \geq 10^{5}$ K) gas out to $R_{200}$, while low-$z$ haloes have $5\times$ more hot gas than cool gas.  The low-$z$ CGM phases are more sorted by radius than for haloes at high $z$, with the cool phase being larger in the inner 50 kpc, and the hot phase dominating at larger radii. [Fig. \ref{fig:massradial}, \ref{fig:masscomp}] 
    
    \item The high-$z$ ISM has $1.5\times$ the total metal content of the high-$z$ CGM, while the low-$z$ CGM has $2.6\times$ the metal content of the low-$z$ ISM.  The high-$z$ hot CGM contains 60\% more metals than the cool CGM, while this reverses at low $z$ with the cool CGM having 35\% more metals than the hot CGM content. [Fig. \ref{fig:metalradial}]
    
    \item The metals are evenly distributed between the hot and cool phases throughout the high-$z$ CGM. At low $z$, the cool metals dominate the interior and the hot metals are more prevalent at larger radii. Cool metallicities increase from about $0.1 \Zsolar$ to $\Zsolar$ from high to low $z$ indicating much of the cool gas is pristine accretion at high $z$ and recycling gas at low $z$.  Hot metals have less scatter and intermediate metallicities that change less across time, which is a signature of their thermal feedback-driven origins using the EAGLE prescription. [Fig. \ref{fig:metalcomp}, \ref{fig:metallicity}]
    
    \item Hot gas shows substantial outflows at high $z$, which stands in contrast to the cool gas that is primarily accreting with the highest inflow velocities being in the outer halo where the gas is relatively pristine.  Low-$z$ radial velocities are much lower with only inner hot gas showing a net outflow, and cool gas accreting at a much lower rate.  [Fig. \ref{fig:velradmass}, \ref{fig:outflowfraction}]
    
    \item Hot metal-enriched ($Z \geq 0.1 \Zsolar$) gas shows larger outflow velocities than all hot gas at both epochs.  Cool metals, like all cool gas, are primarily inflowing at low $z$.  High-$z$ cool metals indicate less inflows than all cool gas, and their kinematics show proportionally more outflows when a higher $Z$ threshold is applied.  The high-$z$ cool CGM has different origins, which depend sensitively on the baryon or metal tracer used. [Fig. \ref{fig:velradmetal}, \ref{fig:outflowfraction}]
    
    \item The cool low-$z$ CGM shows a net positive rotation out to $0.2 R_{200}$, indicating disc-like CGM structures extending out 40 kpc around $L^*$ galaxies.  Hot gas at low $z$ also shows substantially net positive rotation, but no preferred rotation beyond $\approx 50$ kpc.  These low-$z$ hot haloes have been shown to be supported primarily by tangential velocities in the inner CGM and by the thermal pressure gradient in the outer CGM \citep{opp18b}, but no such dynamical stability (i.e. hydrostatic equilibrium) applies to the high-$z$ hot CGM, which is primarily outflowing. [Fig. \ref{fig:veltanmass}]
    
    \item The angular momentum spin parameter of the CGM is substantially higher than that of the dark matter at both epochs. The average hot CGM spin parameters are 66\% of the cool CGM spin parameters at high $z$ and 87\% at low $z$.  Owing to a greater hot CGM mass at low $z$, the total angular momentum in the low-$z$ hot phase is several times that of the cool phase. [Fig. \ref{fig:spinparams}]
    
    \item Angular momentum vectors are well-aligned between the cool and hot CGM at both epochs.  The CGM angular momentum is substantially less well-aligned with the stellar disc, which may result from gas in the outer CGM being dynamically disconnected from the inner CGM.  This does not contradict the existence of the low-$z$ cool CGM often exhibiting co-rotation with the stellar disc. [Fig. \ref{fig:veltanmass}, \ref{fig:spinangles}]
    
    \item Hot halo profiles have flatter density profiles at low $z$ than at high $z$.  High-$z$ hot haloes are hotter and significantly higher pressure than their low-$z$ counterparts.  High-$z$ entropy profiles are rising through most of the halo, while low-$z$ profiles are more isentropic, although with interior positive slopes where hot winds are outflowing. [Fig. \ref{fig:hotprofiles}]
    
\end{enumerate}

Our next paper in this series (Lonardi et al. in prep.) will show that these zoom haloes reproduce key metal absorption line strengths around both star-forming $z\approx 2-3$ and $z\approx 0$ galaxies.  While the typical observed column densities do not change much between these two epochs, our main conclusions here show that these two sets of haloes are physically and dynamically distinct.  Just as galaxies evolve significantly over 10 Gyr, the CGM also evolves, and it will be crucial to identify observational measures that differentiate high-$z$ and low-$z$ gaseous haloes.  

\section*{acknowledgements}

The authors are grateful for valuable discussions with Ryan Horton and Peter Mitchell. Support for EH was provided by the Undergraduate Research Opportunities Program at the University of Colorado Boulder.  BDO was supported through the NASA ATP grant NNX16AB31G and NASA {\it Hubble} grant HST-AR-14308.  RAC is a Royal Society University Research Fellow. AJR was supported by a CO-FUND/Durham Junior Research Fellowship under EU grant 609412; and by the Science and Technology Facilities Council [ST/P000541/1]. This work used the DiRAC@Durham facility managed by the Institute for Computational Cosmology on behalf of the STFC DiRAC HPC Facility (http://www.dirac.ac.uk).  The equipment was funded by BEIS capital funding via STFC capital grants ST/K00042X/1, ST/P002293/1, ST/R002371/1 and ST/S002502/1, Durham University and STFC operations grant ST/R000832/1. DiRAC is part of the National e-Infrastructure.   %[BDO-XXX-CU-HPC]?  

\bibliographystyle{mnras}
\bibliography{EvolvingCGM_2nd_Version.bib}

\end{document}